%% file: main.tex
\documentclass{article}

\PassOptionsToPackage{numbers, compress, sort}{natbib}

\usepackage{amsmath}
\usepackage{amssymb}
\usepackage{mathtools}
\usepackage{amsthm}
\usepackage{array}
\usepackage{multirow}
\usepackage{placeins}

\usepackage[final]{neurips_2024}

\usepackage[utf8]{inputenc} 
\usepackage[T1]{fontenc}    
\usepackage{hyperref}       
\usepackage{url}            
\usepackage{booktabs}       
\usepackage{amsfonts}       
\usepackage{nicefrac}       
\usepackage{microtype}      
\usepackage{xcolor}         
\usepackage{graphicx}
\usepackage{svg}
\usepackage{cleveref}
\usepackage{xcolor}
\usepackage{wrapfig,lipsum,booktabs}
\usepackage{algorithm}
\usepackage{algorithmic}
\usepackage{multicol}
\usepackage{subcaption}
\usepackage{enumitem}
\hypersetup{
    colorlinks=true,
    citecolor=gray,
    }
\usepackage{epigraph}
\usepackage[all=normal, mathspacing=normal,mathdisplays=normal, floats=tight, paragraphs=tight, lists=normal]{savetrees}
\definecolor{lightgreen}{RGB}{144, 238, 144}
\newcommand{\highlightgreen}[1]{\colorbox{lime}{$\displaystyle#1$}}
\newcommand{\highlightblue}[1]{\colorbox{cyan}{$\displaystyle#1$}}
\newtheorem{proposition}{Proposition}
\title{Absorb \& Escape: Overcoming Single Model Limitations in Generating Genomic Sequences}

\author{
  \textbf{Zehui Li\textsuperscript{1,*}, Yuhao Ni\textsuperscript{1}, Guoxuan Xia\textsuperscript{1}, William Beardall\textsuperscript{1},} \\ 
  \textbf{Akashaditya Das\textsuperscript{1}, Guy-Bart Stan\textsuperscript{1,*}, Yiren Zhao\textsuperscript{1,*}} \\
  \textsuperscript{1}Imperial College London, \texttt{\{zehui.li22, harry.ni21, g.xia21, william.beardall15,} \\ 
  \texttt{akashaditya.das13, g.stan, a.zhao\}@imperial.ac.uk} \\
  \textsuperscript{*}Correspondence: \texttt{\{zehui.li22, g.stan, a.zhao\}@imperial.ac.uk}
}

\begin{document}

\maketitle

\begin{abstract}
 Recent advances in immunology and synthetic biology have accelerated the development of deep generative methods for DNA sequence design. Two dominant approaches in this field are AutoRegressive (AR) models and Diffusion Models (DMs). However, genomic sequences are functionally heterogeneous, consisting of multiple connected regions (e.g., Promoter Regions, Exons, and Introns) where elements within each region come from the same probability distribution, but the overall sequence is non-homogeneous. This heterogeneous nature presents challenges for a single model to accurately generate genomic sequences.
 In this paper, we analyze the properties of AR models and DMs in heterogeneous genomic sequence generation, pointing out crucial limitations in both methods: (i) AR models capture the underlying distribution of data by factorizing and learning the transition probability but fail to capture the global property of DNA sequences. (ii) DMs learn to recover the global distribution but tend to produce errors at the base pair level. 
 To overcome the limitations of both approaches, we propose a post-training sampling method, termed \textbf{A}bsorb \textbf{\&} \textbf{E}scape (\textsf{A\&E}) to perform compositional generation from AR models and DMs. This approach starts with samples generated by DMs and refines the sample quality using an AR model through the alternation of the Absorb and Escape steps. 
 To assess the quality of generated sequences, we conduct extensive experiments on 15 species for conditional and unconditional DNA generation. The experiment results from motif distribution, diversity checks, and genome integration tests unequivocally show that \textsf{A\&E} outperforms state-of-the-art AR models and DMs in genomic sequence generation.  \textsf{A\&E} does not suffer from the slowness of traditional MCMC to sample from composed distributions with Energy-Based Models whilst it obtains higher quality samples than single models. Our research sheds light on the limitations of current single-model approaches in DNA generation and provides a simple but effective solution for heterogeneous sequence generation. Code is available at the Github Repo\footnote{https://github.com/Zehui127/Absorb-Escape}.

\end{abstract}

\input{sections/introduction}

\input{sections/preliminary}
\input{sections/post-intro}

\input{sections/method}
\input{sections/experiment}
\input{sections/conclusion}

\begin{ack}
This research project has benefitted from the Microsoft Accelerate Foundation Models Research (AFMR) grant program. Zehui Li is grateful to his family—Jinghai Li, Huiqing Wang, and Yiqiu Sun for their support. 
\end{ack}

\bibliographystyle{ieee_fullname_natbib}
\bibliography{reference}
\newpage
\input{Appendix/appendix}

\FloatBarrier
\input{sections/checklist}
\FloatBarrier
\newpage
\end{document}

%% file: sections/introduction.tex
\section{Introduction}
\begin{figure}[h]
    \centering
    \begin{subfigure}[b]{0.37\textwidth}
        \centering
        \includegraphics[width=\linewidth]{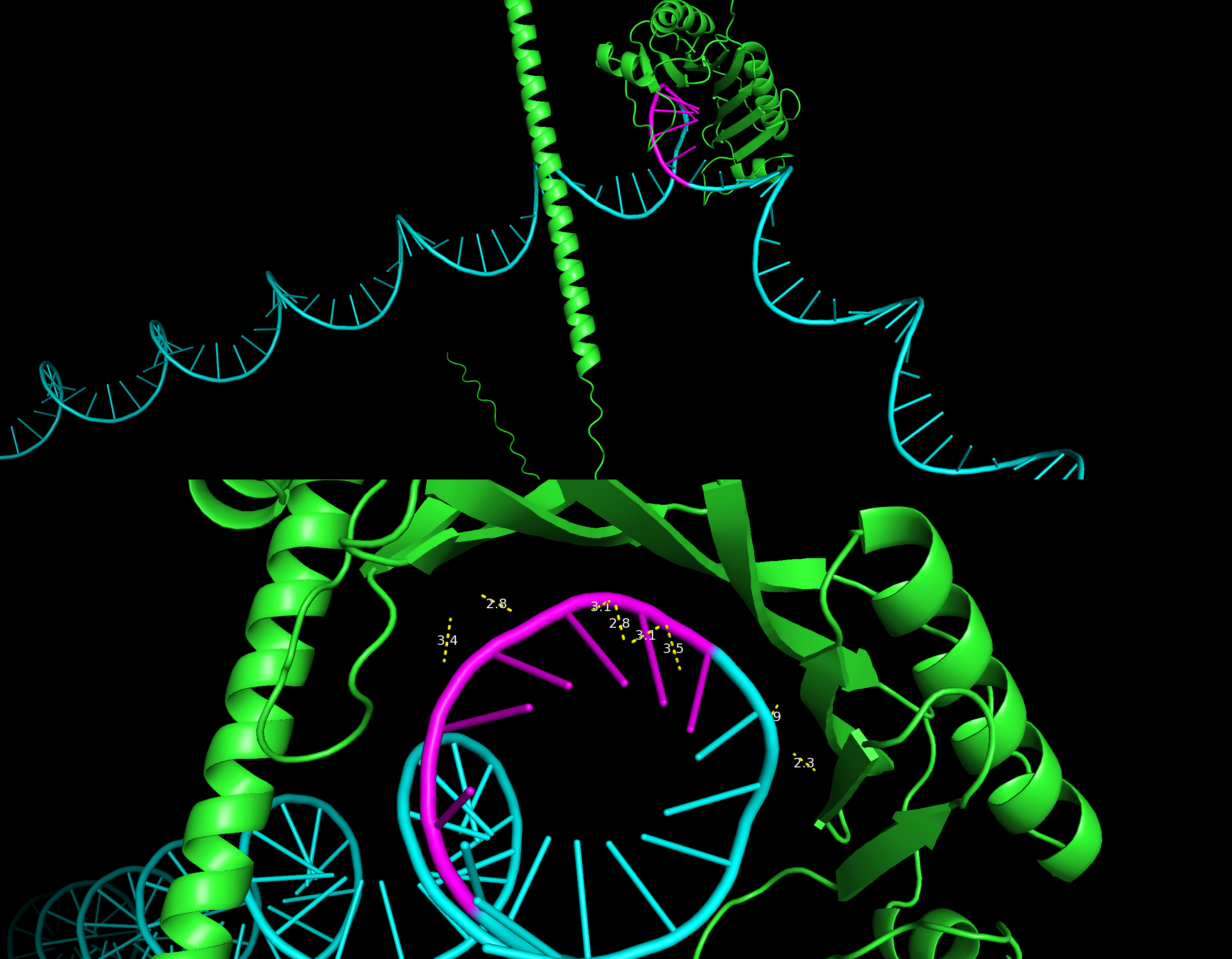}
        \caption{\textbf{A DNA generated by \textsf{A\&E}, interacting with TATA-binding protein.} The DNA sequences highlighted in \textcolor{magenta}{magenta} are the TATA-box motif. The confirmation is predicted by AlphaFold 3~\citep{abramson2024accurate}: the DNA bends at the TATA-box position.}
        \label{fig:tata-demo}
    \end{subfigure}
    \hfill
    \begin{subfigure}[b]{0.59\textwidth}
        \centering
        \includegraphics[width=\linewidth]{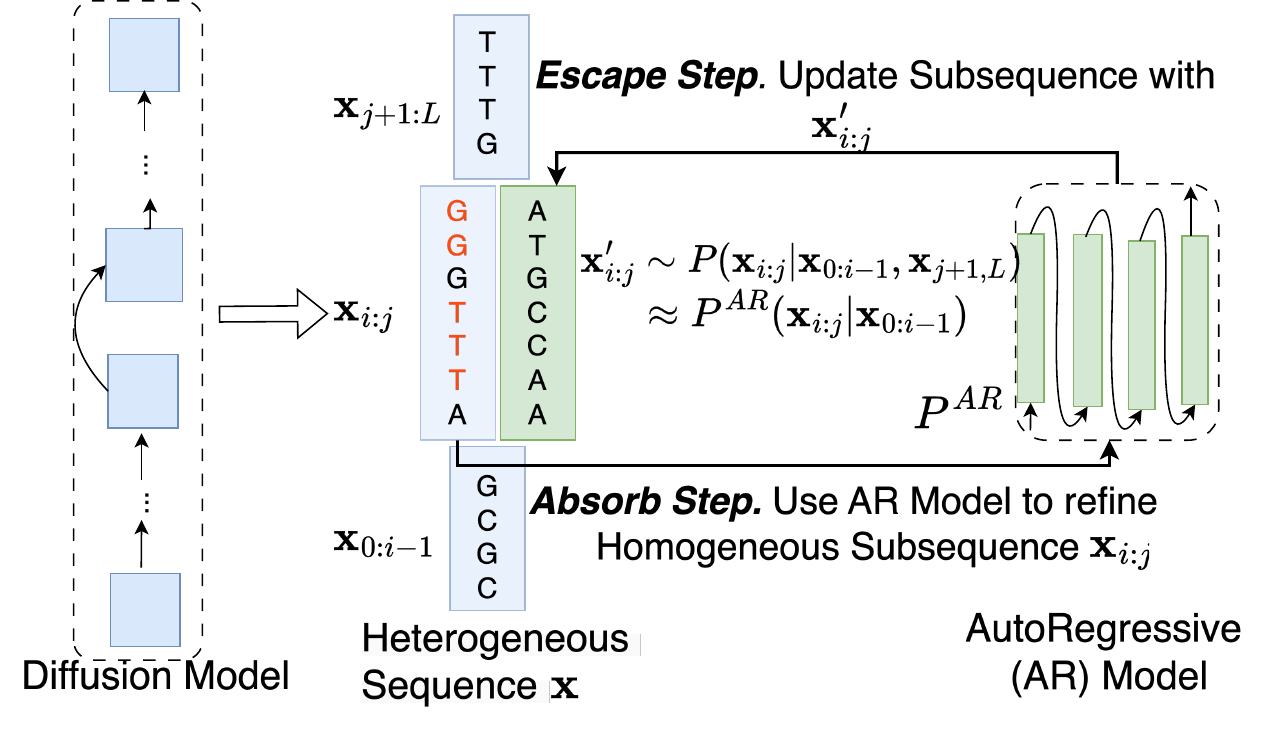}
        \caption{The proposed framework Fast \textbf{A}bsorb \textbf{\&} \textbf{E}scape (\textsf{Fast A\&E}): The DM and AR models jointly optimize a given sequence by alternating between the A-step and the E-step. See \Cref{sec:method} for a detailed explanation.}
        \label{fig:framework}
    \end{subfigure}
    \caption{(a) Generated DNA interacting with TATA-binding protein. (b) Proposed \textsf{A\&E} framework.}
    \label{fig:combined}
    \vspace{-4mm}
\end{figure}

DNA sequences, as the blueprint of life, encode proteins and RNAs and directly interact with these molecules to regulate biological activities within cells. The success of deep generative models in image~\citep{rombach2022high}, text~\citep{radford2019language}, and protein design~\citep{watson2023novo} has drawn the attention of deep learning researchers to the problem of DNA design, i.e. applying these models to genomic sequence generation~\citep{avdeyev2023dirichlet, zrimec2022controlling, wang2020synthetic}. However, one rarely explored issue is how well existing methods can handle the unique property of DNA sequences: heterogeneity. DNA sequences are highly heterogeneous, consisting of multiple connected functional regions (e.g. Promoter Regions, Exons, and Introns) in sequential order. While elements within each functional region might be homogeneous (coming from the same distribution), the overall sequence is non-homogeneous. This heterogeneity, along with the discrete nature of genomic sequences, poses challenges to popular deep generative methods.
 
\paragraph{Limitations of Existing Single-Model Approaches in Generating Genomic Sequences}
AutoRegressive Models (AR)~\citep{nguyen2024hyenadna,dalla2023nucleotide,ji2021dnabert} are one of the most dominant approaches for discrete sequence generation. To model the data distribution of a sequence \( x \) of length \( T \), the probability of \( x \) is factorized as:

\begin{equation} \label{eq:1}
    p^{AR}(\mathbf{x}) = \prod_{i=1}^{T}p_\theta(x_i | x_1, x_2, \dots, x_{i-1}).
\end{equation}

An issue arises when modeling heterogeneous data, where the value of \( \theta \) may vary significantly from one segment to another. Additionally, AR models assume a dependency between the current element and previous elements; this assumption may not hold true for heterogeneous sequences, potentially hindering the learning process (see \Cref{sec:toy-example} for details).

On the other hand, Diffusion Models (DMs), initially proposed by ~\citep{sohl2015deep}, have been dominant in image generation. In the probabilistic denoising view~\citep{ho2020denoising}, DMs gradually add noise to the input data \( x_0 \), and a reverse diffusion (generative) process is trained to gradually remove the noise from the perturbed data \( x_t \). DMs directly model the data distribution without AutoRegressive factorization, thereby avoiding the issues associated with AR models. However, it has been shown that DMs are less competent than AR models for discrete data generation~\citep{zhang2024planner,li2022diffusion}. When it comes to modeling heterogeneous genomic sequences, it remains unclear how the performance of DMs compares to AR models within each homogeneous segment.
\paragraph{Model Composition As a Solution}
Balancing the ability of generative algorithms to capture both local and global properties of the data distribution is central to the problem. An obvious solution could be to combine these two types of models and perform generation using the composed models. However, this typically requires converting these two models into an Energy-Based Model and then sampling using Markov Chain Monte Carlo (MCMC), which can be inherently slow due to the sampling nature of the algorithm~\citep{du2024compositional}, and the potential long inference time of individual models. With the goal of accurate and efficient DNA generation, we aim to investigate two key questions in this work: 
(i) How well does a single AR model or DM perform in DNA generation, given the heterogeneous nature of genomic sequences? 
(ii) Is there an efficient algorithm to combine the benefits of AR models and DMs, outperforming a single model? In answering these two questions, \textit{our contribution is three-fold}:

\begin{enumerate}[label=(\alph*)]
\item We study the properties of AR models and DMs in heterogeneous sequence generation through theoretical and empirical analysis (\Cref{sec:toy-example}).
\item We design the theoretical framework \textbf{A}bsorb \textbf{\&} \textbf{E}scape (\textsf{A\&E}) to sample from the compositional distribution of an AR model and a DM (\Cref{sec:theory-ae}). Furthermore, as shown in \Cref{fig:framework}, we propose an efficient post-training sampling algorithm termed \textsf{Fast A\&E} to sample from the composed model, requiring at most one forward pass through the pretrained DM and AR model (\Cref{sec:ae-implementation}).
\item We design a comprehensive evaluation workflow for DNA generation, assessing the sequence composition, diversity, and functional properties of generated genomic sequences. Extensive experiments (15 species, 6 recent DMs, 1 AR model, and 3 types of evaluations) reveal: 
  1) the limitations of existing models in DNA generation (\Cref{sec:main-result}), and 
  2) that the proposed algorithm \textsf{Fast A\&E} consistently outperforms state-of-the-art models as measured by motif distribution and functional property similarity to natural DNA sequences (\Cref{sec:ae-implementation,sec:main-result}).
\end{enumerate}



%% file: sections/preliminary.tex
\section{Preliminaries and Related Work}
\subsection{Problem Formulation: DNA Generation}
DNA generation aims to produce synthetic sequences that functionally approximate real DNA sequences. Formally, let \( \mathbb{N}_4 = \{1, 2, 3, 4\} \), where each element represents one of the four nucleotides: adenine (A), thymine (T), guanine (G), and cytosine (C). A DNA sequence of length \( L \) can be represented as \( \mathbf{x} \in \mathbb{N}^L_4 \), with each element/nucleotide denoted by \( \mathbf{x}_1, \mathbf{x}_2, \cdots, \mathbf{x}_L \).

\textbf{Unconditional Generation:} Given a dataset of real-world DNA sequences \(\mathcal{X} = \{\mathbf{x}^{(n)}\}_{n=1}^N\) collected from some distribution \( p(\mathbf{x}) \), where each sequence \(\mathbf{x}^{(n)} \in \mathbb{N}^L_4\) represents a chain of nucleotides, the objective is to develop a generative model \( p_{\boldsymbol\theta}(\mathbf{x}) \) of the data distribution \( p(\mathbf{x}) \) from which we can sample novel sequences \(\tilde{\mathbf{x}} \sim p_{\boldsymbol\theta}(\mathbf{x}) \). These sequences should be structured arrangements of A, T, G, and C, reflecting the complex patterns found in actual DNA. Earlier works applying Generative Adversarial Networks (GANs)~\citep{goodfellow2014generative} to generate protein-encoding sequences~\citep{gupta2019feedback} and functional elements~\citep{taskiran2023cell,zrimec2022controlling,wang2020synthetic} fall into this category.

\textbf{Conditional Generation:} In this task, the dataset of DNA sequences \(\mathcal{X} = \{\mathbf{x}^{(n)}, c^{(n)}\}_{n=1}^N\) is sampled from the joint distribution \( p(\mathbf{x}, c) \), where \( c \) represents the condition associated with each sequence. The objective is to develop a model \( p_{\boldsymbol\theta}(\mathbf{x}|c) \) that generates new DNA sequences \(\tilde{\mathbf{x}}\) given condition \( c \). Recently, a discrete diffusion model DDSM~\citep{avdeyev2023dirichlet} and an AutoRegressive model RegML~\citep{lal2024reglm} have used expression level as the condition, while DNADiffusion~\citep{pinello2024dna} has used cell type as the condition.

\subsection{Homogeneous vs. Heterogeneous Sequences}

\textbf{Homogeneous Generation Process}
In the context of sequence generation, a homogeneous Markov Chain is characterized by constant probabilistic rules for generating the sequence at each time step \( t \). More generally, a process is defined as \textbf{homogeneous} if the transition probabilities are independent of time \( t \). This means there exists a constant \( P_{\mathbf{c},j} \) such that:
\begin{equation}
 P_{\mathbf{c},j} = \Pr[\mathbf{x}_t = j \mid \mathbf{x}_{1:t-1} = \mathbf{c}]    
\end{equation}
holds for all times \( t \), where \( P_{\mathbf{c},j} \) is a constant, and \( \mathbf{c} \) represents a specific sequence of past values, i.e., \( \mathbf{c} = (c_1, c_2, \ldots, c_{t-1}) \).

\textbf{Heterogeneous Generation Process}
Assuming homogeneous properties simplifies modeling but can be overly restrictive for certain modalities, leading to inaccuracies. For example, DNA sequences consist of various functionally distinct regions, such as promoters, enhancers, regulatory regions, and protein-coding regions scattered across the genome~\citep{eraslan2019deep}. Each region may be assumed to be homogeneous, but the overall sequence is non-homogeneous due to the differing properties of these elements. 

For sequences like DNA, which consist of locally homogeneous segments, we define them as \textbf{heterogeneous sequences}. Formally, a heterogeneous sequence is defined as follows: Suppose a sequence \( \mathbf{x} \) is divided into segments \( \mathcal{S}_1, \mathcal{S}_2, \ldots, \mathcal{S}_m \), where each segment \( \mathcal{S}_i \) is homogeneous. For each segment \( \mathcal{S}_i = (\mathbf{x}_t, \mathbf{x}_{t-1}, \mathbf{x}_{t-2}, \ldots, \mathbf{x}_{t-k}) \), there exists a constant \( P_{\mathbf{c}_i, j} \) such that:
\begin{equation}
P_{\mathbf{c}_i, j} = \Pr[\mathbf{x}_t = j \mid \mathbf{x}_{t-k:t-1} = \mathbf{c}_{i}]
\end{equation}
holds for all times \( t \) within segment \( \mathcal{S}_i \), where \( P_{\mathbf{c}_i, j} \) is a constant, and \( \mathbf{c}_i = (c_{i, t-1}, c_{i, t-2}, \ldots, c_{i, t-k}) \) represents values characterizing the history within that segment. While segment \( \mathcal{S}_i \) is homogeneous, the entire sequence is non-homogeneous due to the varying properties across different segments.

%% file: sections/post-intro.tex
\section{Single Model Limitations in Heterogeneous Sequence Generation}
\label{sec:toy-example}

\begin{figure}
	\centering
	\includegraphics[width=1\linewidth]{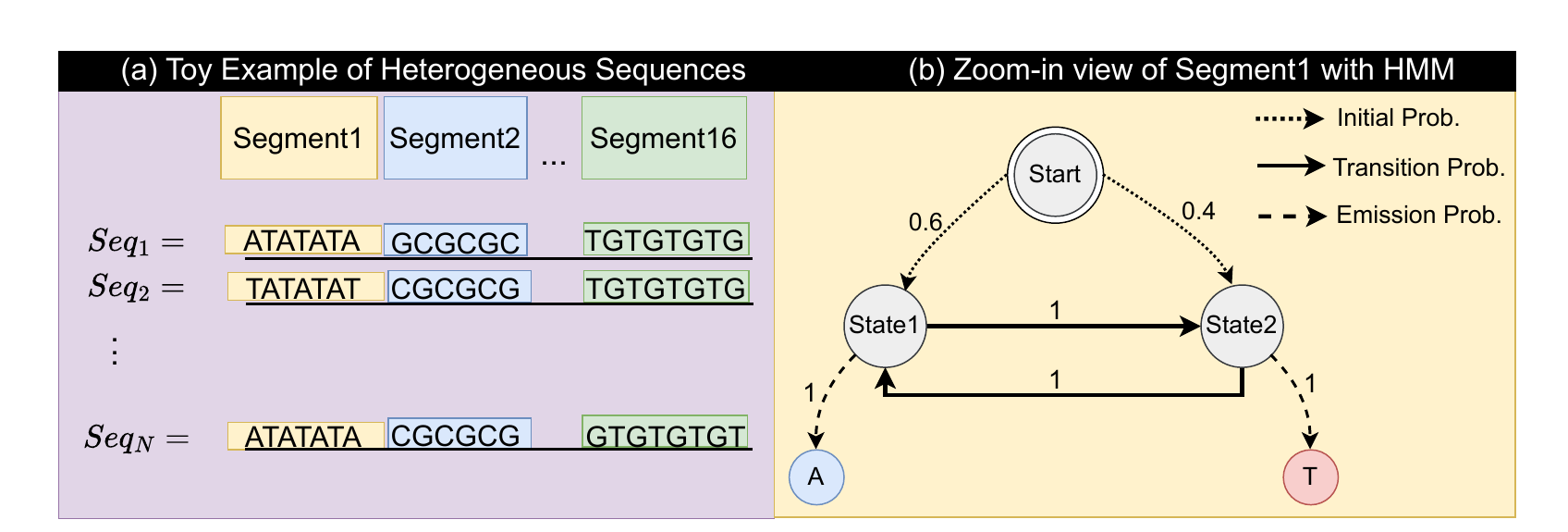}
	\caption{A toy example with heterogeneous sequences: (a) The overall training set consists of $N=50,000$ heterogeneous sequences, where each sequence further consists of 16 homogeneous segments. We apply an autoregressive and a diffusion model to learn the underlying distribution. (b) Within each segment, the sequences are generated with a simple Hidden Markov Model (HMM), with deterministic transition probability and emission probability. }
	\label{fig:toy_hmm}
 \vspace{-4mm}
\end{figure}

How powerful are AR models and DMs in modelling heterogeneous sequences? We first provide a theoretical analysis, and then perform experiments on synthetic sequences to validate our assumption. 

\textbf{AutoRegressive (AR) Models} Suppose a heterogeneous sequence  $\mathbf{x}$ consist of two homogeneous segments of length k, then $\mathbf{x}= \{\{x_1,x_2,\cdots,x_k\},\{x_{k+1},x_{k+2},\cdots,x_{2k}\}\}$. AR models factorize $p(\mathbf x)$ into conditional probability in \cref{eq:exp-factor}; consider the case where the true factorisation of $p(x)$ follows \cref{eq:exp-true}. 

\begin{align}    
    p^{AR}(\mathbf{x}) = 
   p_\theta(x_1)p_\theta(x_2|x_1)\cdots p_\theta(x_k|\mathbf{x}_{1:k-1})
    \cdot p_\theta(x_{k+1}|\mathbf{x}_{1:k})p_\theta(x_{k+2}|\mathbf{x}_{1:k+1})\cdots p_\theta(x_{2k}|\mathbf{x}_{1:2k-1})
\label{eq:exp-factor}
\end{align}

\begin{align}
   p^{data}(\mathbf{x}) =  \underbrace{p_1(x_1)p_1(x_2|x_1)\cdots p_1(x_k|\mathbf{x}_{1:k-1})}_{\text{Segment 1}} \cdot  \underbrace{p_2(x_{k+1})p_2(x_{k+2}|\mathbf{x}_{k+1})\cdots p_2(x_{2k}|\mathbf{x}_{k+1:2k-1})}_{\text{Segment 2}}
\label{eq:exp-true}
\end{align}

AR factorisation allows the accurate modelling of the first homogeneous segment; however, it may struggle to disassociate the elements of the second segment from the first segment. More precisely, sufficient data is needed for AR model to learn that $p_\theta(x_{k+1}), p_\theta(x_{k+2}), \cdots, p_\theta(x_{2k})$ should be independent to the elements $\{x_1, x_2, \cdots, x_{k}\}$ in the first segments. Secondly, when the context length of the AR model is shorter than the sequence length $2k$, it could struggle to capture the difference between $p_1$ and $p_2$ with a single set of parameters $\theta$.

\textbf{Diffusion Models (DMs)} On the other hand, DMs estimate the overall probability distribution $p(\mathbf{x})$ without factorization. The elements of $\mathbf{x}$ are usually generated in parallel. Thus, they do not suffer from the conditional dependence assumption. However, the removal of the conditional dependence assumption may also decrease the accuracy of generation within each homogeneous segment compared to AR models, as DMs do not explicitly consider previous elements. 

\subsection{A Toy Example}
To evaluate the performance of Autoregressive (AR) models and Diffusion Models (DMs) in generating heterogeneous sequences, we consider a toy example with 50,000 heterogeneous sequences $\mathcal{X} = \{\mathbf{x}^{(n)}\}_{n=1}^{50000}$. Each sequence contains 16 segments, as illustrated in \Cref{fig:toy_hmm}(a), and each segment comprises 16 elements, resulting in a total sequence length of 256 ($\mathbf{x} \in \mathbb{N}^{256}_4$). A simple Hidden Markov Model (HMM) is used to generate each segment, as shown in \Cref{fig:toy_hmm}(b), with deterministic transition and emission probabilities that ensure homogeneity within each segment. The emitted tokens differ from one segment to another, mimicking the properties of real DNA sequences. Whilst it is possible to use more complex distributions for each segment, doing so could complicate the evaluation of the generated sequences.
 
\textbf{Evaluation} Under our toy HMM setup, a generative model could make two types of mistakes within each segment: \textbf{1) Illegal Start Token}: The generated sequence starts with a token which has zero emission probability. E.g. in \Cref{fig:toy_hmm}(b), the starting token could only be $\{\textcolor{lightgreen}{A,T}\}$. \{\textcolor{red}{$G,C$}\} at the beginning of the sequence are classified as illegal start tokens. \textbf{2) Incorrect Transition}: The generated sequence contains tokens with zero transition probability. E.g. in \Cref{fig:toy_hmm}(b) given the start of the sequence is (\textcolor{lightgreen}{$A,T,A,T$}), the next token must be \textcolor{lightgreen}{$A$}, any other tokens such as \{\textcolor{red}{$T,G,C$}\} are classified as incorrect transitions. We use the number of incorrect tokens as the metric for evaluation.

\begin{wraptable}[10]{r}{0.65\textwidth}  
      \vspace{-5mm}
      \small
      \caption{{\bf Number of Incorret Tokens on Synthetic Dataset}. The performance metrics used are the number of Illegal Start (IS) Tokens and Incorrect Transition (IT) Tokens. Note that there are a total of $4,000 \times 256 = 1024,000$ tokens.
      }
      \label{tab:toy-example}
    \vspace{2mm}
    \centering
    \setlength{\tabcolsep}{4pt} 
\begin{tabular}{lcc}
    \toprule
    & \textsc{HyenaDNA} & \textsc{DiscDiff} \\
    \midrule
    \textsc{\# IS Tokens $\downarrow$} & 812 & \textbf{0} 
    \\
    \midrule
    \textsc{\# IT Tokens $\downarrow$} & \textbf{3,586} & 110,192 \\
    \bottomrule
\end{tabular}

\end{wraptable}

\textbf{Experiment} We use HyenaDNA~\citep{nguyen2024hyenadna,lal2024reglm} as the representative autoregressive (AR) model. For the diffusion model, we develop a simple latent Discrete Diffusion model termed DiscDiff. It resembles the design of StableDiffusion~\citep{rombach2022high}, a latent diffusion model for image generation. DiscDiff consists of a CNN-based Variational Encoder-decoder, trained with cross entropy, to map the discrete DNA data into a latent space, and a standard 2-D UNet as the denoising network (detailed in \cref{app:discdiff}). 
The training dataset consists of $\mathcal{X} = \{\mathbf{x}^{(n)}\}_{n=1}^{50,000}$. For detailed training procedures see \Cref{app:toy-exp-setup}. We generate 4,000 sequences from each model and present the evaluation results in \Cref{tab:toy-example}. As expected, the diffusion model DiscDiff makes fewer errors regarding Illegal Start (IS) tokens but tends to generate more Incorrect Transition (IT) tokens. Conversely, while the AR model HyenaDNA generates some IS token errors, it produces significantly fewer IT token errors. This motivates the question: \textit{can we combine the strengths of both algorithms to achieve better sequence generation?}

%% file: sections/method.tex
\section{Method}
\label{sec:method}
\begin{table}[h!]
\begin{minipage}[t]{0.48\textwidth}
\begin{algorithm}[H]
\caption{Absorb \& Escape Algorithm}
\begin{algorithmic}[1]
\REQUIRE Pretrained AutoRegressive model $p^{AR}_\theta(\mathbf{x})$ and pretrained Diffusion Model $p^{DM}_\beta(\mathbf{x})$
\STATE Initialize $\tilde{\mathbf{x}}^0 \sim p^{DM}_\beta(\mathbf{x})$
\STATE Set $t = 0$
\STATE Assume $\mathbf{x} = \{\mathbf{s}_1, \mathbf{s}_2, \ldots, \mathbf{s}_n\}$, where each $\mathbf{s}_k = \{\mathbf{x}_i, \mathbf{x}_{i+1}, \cdots, \mathbf{x}_j\}$ is a segment
\WHILE{not converged}
\STATE Sample a segment $\mathcal{S} \in \{\mathbf{s}_1, \mathbf{s}_2, \ldots, \mathbf{s}_n\}$
    \STATE Set $i = \text{start index of } \mathbf{s}_k$, $j = \text{end index of } \mathbf{s}_k$
    \STATE   \highlightgreen{\textbf{Absorb step:}}
    \STATE $\tilde{\mathbf{x}}_{i:j}' \sim p(\mathbf{x}_{i:j}|\mathbf{x}_{0:i-1}, \mathbf{x}_{j+1:L}) \approx p^{AR}_\theta(\mathbf{x}_{i:j}|\mathbf{x}_{0:i-1})$ \textit{//Refine segment $\mathbf{s}_k$ using the AR model}
    \STATE  \highlightblue{\textbf{Escape step:}}
    \STATE $\tilde{\mathbf{x}}_{i:j}^t = \tilde{\mathbf{x}}_{i:j}'$ \textit{//Update $\tilde{\mathbf{x}}^t$}
    \STATE Increment $t = t + 1$
\ENDWHILE
\STATE \textbf{Output:} Final sample $\tilde{\mathbf{x}}^t$ with improved quality
\end{algorithmic}
\label{alg:ae}
\end{algorithm}

\end{minipage}
\hfill
\begin{minipage}[t]{0.48\textwidth}
\begin{algorithm}[H]
\caption{Fast Absorb \& Escape Algorithm}
\begin{algorithmic}[1]
\REQUIRE Absorb Threshold $T_{Absorb}$, Pretrained AutoRegressive model $p^{AR}_\theta(\mathbf{x})$ and pretrained Diffusion Model $p^{DM}_\beta(\mathbf{x})$
\STATE Initialize $\tilde{\mathbf{x}}^0 \sim p^{DM}_\beta(\mathbf{x})$
\FOR{$i$ in $len(\mathbf{\tilde{x}})$}
    \IF{$p^{DM}<T_{Absorb}$}
        \STATE \highlightgreen{\textbf{Absorb step:}}
        \STATE j = i+1
        \STATE $\tilde{\mathbf{x}}_{j}' \sim p^{AR}_\theta(\mathbf{x}_{j}|\mathbf{x}_{0:i})$
        \WHILE{$p^{AR}(\tilde{\mathbf{x}}_j')>p^{DM}(\tilde{\mathbf{x}}_j)$}
            \STATE Increment $j=j+1$
            \STATE $\tilde{\mathbf{x}}_{j}' \sim p^{AR}_\theta(\mathbf{x}_{j}|\mathbf{x}_{0:i},\mathbf{x}_{i:j-1})$ \textit{//Refine Inaccurate region of the sequence token by token}
        \ENDWHILE
            \STATE \highlightblue{\textbf{Escape step:}}
    \STATE $\tilde{\mathbf{x}}_{i:j} = \tilde{\mathbf{x}}_{i:j}'$ \textit{//Update $\tilde{\mathbf{x}}$}
    \STATE Increment i = i + j
    \ENDIF
\ENDFOR
\STATE \textbf{Output:} $\tilde{\mathbf{x}}$ with improved quality
\end{algorithmic}
\label{alg:ae-fast}
\end{algorithm}
\end{minipage}
\end{table}

\subsection{The Absorb \& Escape Framework}
\label{sec:theory-ae}
Given a pretrained AutoRegressive model $p^{AR}_\theta(\mathbf{x})$ and a Diffusion Model $p^{DM}_\beta(\mathbf{x})$, we aim to generate a higher quality example $\tilde{\mathbf{x}}$ from the composed distribution $p^{C}_{\theta,\beta}(\mathbf{x})= p^{AR}_\theta(\mathbf{x}) \circ p^{DM}_\beta(\mathbf{x})$. However, directly computing $p^{C}_{\theta,\beta}(\mathbf{x})$ is generally intractable, as both the autoregressive factorizations from $p^{AR}$ and score functions from $p^{DM}$ are not directly composable~\citep{du2024compositional,du2023reduce}. We propose the \textbf{A}bsorb \textbf{\&} \textbf{E}scape (\textsf{A\&E}) framework, as shown in \Cref{alg:ae}, to efficiently sample from $p^{C}_{\theta,\beta}(\mathbf{x})$.

\paragraph{Absorb-Step}
Inspired by Gibbs sampling~\citep{gelfand2000gibbs}, which iteratively refines each dimension of a single sample and moves to higher density areas, our algorithm starts with a sequence $\mathbf{x}^0 \sim p^{DM}(\mathbf{x})$, generated by the diffusion model and then refines the samples through the Absorb step and Escape step. By exploiting the heterogeneous nature of the sequence, we assume that \(\mathbf{x}^0\) can be factorized into multiple segments \(\{\mathbf{s}_1, \mathbf{s}_2, \ldots, \mathbf{s}_n\}\). For each segment \(\mathbf{s}_k\), we set \(i\) and \(j\) as the start and end indices, respectively. During the Absorb step, we sample a subset of segments $\mathcal{S} \subseteq \{\mathbf{s}_1, \mathbf{s}_2, \ldots, \mathbf{s}_n\}$ and refine each segment \(\mathbf{s}_k\) by sampling \(\tilde{\mathbf{x}}_{i:j}' \sim p(\mathbf{x}_{i:j}|\mathbf{x}_{0:i-1}, \mathbf{x}_{j+1:L}) \approx p^{AR}_\theta(\mathbf{x}_{i:j}|\mathbf{x}_{0:i-1})\), using the autoregressive model to approximate the conditional probability.

\paragraph{Escape-Step}
After refining the segment in the Absorb step, we proceed with the Escape step where we update the refined segment \(\tilde{\mathbf{x}}_{i:j}^t\) to \(\tilde{\mathbf{x}}_{i:j}'\). This iterative process continues for each selected segment \(\mathbf{s}_k\), with \(t\) incrementing after each update. By leveraging the ability of the diffusion model to capture the overall data distribution and the autoregressive model to refine homogeneous sequences within each segment, our algorithm efficiently improves the quality of the generated samples. The final output \(\tilde{\mathbf{x}}^t\) is hereby closer to the true data distribution $p(\mathbf{x})$ compared to the initial sample \(\tilde{\mathbf{x}}^0\).  A proof for the convergence in \Cref{prop:convergence} is provided in \Cref{app:proof-ab-convergence}.

\begin{proposition}
\label{prop:convergence}
The Absorb \& Escape (\textsf{A\&E}) algorithm converges to the target distribution \(p^{C}_{\theta,\beta}(\mathbf{x}) = p^{AR}_\theta(\mathbf{x}) \circ p^{DM}_\beta(\mathbf{x})\), under the assumptions that both models are properly trained, the segments of \(\mathbf{x}\) are homogeneous, the subset of segments is chosen randomly, and the conditional distribution \(p(\mathbf{x}_{i:j}|\mathbf{x}_{0:i-1}, \mathbf{x}_{j+1:L})\) is accurately approximated by \(p^{AR}_\theta(\mathbf{x}_{i:j}|\mathbf{x}_{0:i-1})\).
\end{proposition}

\subsection{Practical Implementation: Fast A\&E}
\label{sec:ae-implementation}

While \textsf{A\&E} offers a method to sample from a compositional distribution, two practical issues remain unresolved. Firstly, the algorithm may take a considerable amount of time to converge. Secondly, a crucial step in Line 3 of \Cref{alg:ae} involves splitting $\mathbf{x}$ into homogeneous segments $\{\mathbf{s_1}, \mathbf{s_2}, \cdots, \mathbf{s_n}\}$ and then sampling a subset of these segments. Segmentation is straightforward when the boundaries of functional regions of the DNA sequence are known, such as protein-coding regions, exons, or introns, where each region naturally forms a homogeneous segment. However, this information is often unavailable in practice.

To address these challenges, we propose a practical implementation termed \textsf{Fast A\&E}. For generating a sequence $\mathbf{x} \in \mathbb{N}^L_4$, it requires at most $L$ forward passes through the AR model. As shown in \Cref{alg:ae-fast} and \Cref{fig:framework}, \textsf{Fast A\&E} adopts a heuristic-based approach to select segments for refinement. It scans the sequence from left to right, identifying low-quality segments through a thresholding mechanism. Tokens with predicted probabilities smaller than the $T_{absorb}$ threshold trigger the absorb action, while the autoregressive process terminates once the probability of a token generated by the AR model $p^{DM}(\tilde{\mathbf{x}}_j')$ is smaller than that of the diffusion model $p^{AR}(\tilde{\mathbf{x}}_j)$. In this manner, \textsf{Fast A\&E} corrects errors made by the diffusion model with a maximum running time of $O(T_{DM} + T_{AR})$, where $T_{DM}$ and $T_{AR}$ are the times required for generating a single sequence from the diffusion model and autoregressive model, respectively.

%% file: sections/experiment.tex
\section{Experiment}

\subsection{Transcription Profile (TP) conditioned Promoter Design}

We first evaluate \textsf{Fast A\&E} in the task of TP-conditioned promoter design, following the same evaluation procedures and metrics as used by DDSM~\citep{avdeyev2023dirichlet} and Dirichlet Flow Matching (DFM)~\citep{stark2024dirichlet}.

\begin{wraptable}[16]{r}{0.57\textwidth}
    \centering
       \footnotesize
        \caption{\textbf{Evaluation of transcription profile conditioned promoter sequence design.} \textsf{A\&E} achieves the smallest MSE with Language Model and DFM distilled being the AR and DM components.
    }
    \begin{tabular}{l c}
        \toprule
        \textbf{Method} & \textbf{MSE$\downarrow$} \\
        \midrule
        \textbf{Bit Diffusion (bit-encoding)*} & .0414 \\
        \textbf{Bit Diffusion (one-hot encoding)*} & .0395 \\
        \textbf{D3PM-uniform*} & .0375 \\
        \textbf{DDSM*} & .0334 \\
        \textbf{Language Model*} & .0333 \\
        \textbf{Linear FM*} & .0281 \\
        \textbf{Dirichlet FM (DFM)*} & .0269 \\
        \textbf{Dirichlet FM distilled (DFM distilled)*} & .0278 \\
        \midrule
        \textbf{ A\&E (Language Model+Dirichlet FM distilled)} & \textbf{.0262} \\
        \bottomrule
    \end{tabular}
    \label{tab:comparison}
\end{wraptable}

\textbf{Data Format \& Evaluation Metric:} Each data point in this task is represented as a $(\text{DNA}, \text{signal})$ pair, where \textit{signal} corresponds to the CAGE values for a given DNA sequence, providing a quantitative measure of gene expression around the transcription start site (TSS). Both the DNA sequence and the signal have a length of 1024. The goal of this task is to generate DNA sequences conditioned on specified signals. During evaluation, for a given test set data point $(x, c)$, the generated sequence $\tilde{\mathbf{x}}$ and the ground truth sequence are processed through the genomic neural network Sei~\citep{zhou2015predicting}. The performance metric is the mean squared error (MSE) between $Sei(x)$ and $Sei(\tilde{\mathbf{x}})$.

\textbf{Results:} As shown in \Cref{tab:comparison}, we ran \textsf{Fast A\&E} on this task with a default threshold of $T_{\text{absorb}} = 0.85$, using the pretrained model as both the autoregressive (AR) model and the denoising model (DM) component. The evaluation followed the same procedure as described in the DFM repository. The \textsf{Fast A\&E} model, comprising AR and DM components (i.e., the language model and the distilled DFM checkpoints provided in the DFM repository), achieved state-of-the-art results with an MSE of 0.0262 on the test split.

\subsection{Multi-species Promoter Generation}

\subsubsection{Experimental Setup}
\textbf{Dataset Construction} Prior efforts in DNA generation have been constrained by small, single-species datasets~\citep{wang2020synthetic, killoran2017generating}. To better evaluate the capability of various generative algorithms in DNA generation, we construct a dataset with 15 species from the Eukaryotic Promoter Database (EPDnew)\citep{meylan2020epd}. \Cref{table:dataset} compares our EPD dataset with those used in previous studies, including DDSM\citep{avdeyev2023dirichlet}, ExpGAN~\citep{zrimec2022controlling}, and EnhancerDesign~\citep{taskiran2023cell}. The key advantage of EPD is its diversity in both species types and DNA sequence types. Additionally, although the number of sequences in EPD is on a similar scale to that of DDSM, EPD offers greater uniqueness: each sequence corresponds to a unique promoter-gene combination, a guarantee not provided by the other datasets.

\begin{wraptable}[13]{r}{0.6\textwidth}
    \small
    \caption{\textbf{A comparison of DNA generation datasets.} EPD used in this work is significantly larger in size and contains fifteen species. Reg. represents the regulatory regions, and Prot. represents the protein encoding region.}
    \label{table:dataset}
    \vspace{4mm}
    \centering
    \setlength{\tabcolsep}{4pt} 
    \begin{tabular}{@{}llll@{}}
        \toprule
        Dataset & \# DNA & Multi Species  & DNA Regions\\ 
        \midrule
        EPD (Ours) & \textbf{160,000} & \checkmark  & Reg. \& Prot.\\
        DDSM~\cite{avdeyev2023dirichlet} & 100,000 & $\times$  & Reg. \& Prot. \\
        ExpGAN~\cite{zrimec2022controlling} & 4238 & $\times$  & Reg.\\
        EnhancerDesign~\cite{taskiran2023cell} & 7770 & $\times$  & Reg.\\
        \bottomrule
    \end{tabular}
\end{wraptable}

\textbf{Baseline Model} We evaluate the state-of-the-art diffusion models for DNA sequence generation: \textit{DDSM}~\cite{avdeyev2023dirichlet}, \textit{DNADiffusion}~\cite{pinello2024dna}, \textit{DDPM}~\cite{alamdari2023protein, austin2021structured}, and a AR model \textit{Hyena}~\cite{nguyen2024hyenadna, lal2024reglm}. In addition, we implement a VAE with a CNN-based encoder-decoder architecture. Adding UNet as the denoising network to VAE results in another baseline latent diffusion model termed \textit{DiscDiff}. For a fair evaluation, we maximally scale up the denoising networks of each diffusion model to fit into an 40GB NVIDIA A100.  Additionally, we adapt four pretrained \textit{Hyena} models from HuggingFace for comprehensive fine-tuning. The additional details of the network architectures are shown in \Cref{app:baseline}.

\paragraph{Model Training}
All the models are implemented in Pytorch and trained on a NVIDIA A100-PCIE-40GB with a maximum wall time of 48 GPU hours per model; most of the models converged within the given time. Adam optimizer~\citep{da2014method} is used together with the CosineAnnealingLR~\citep{loshchilov2016sgdr} scheduler. The learning rate of each model are detailed in \Cref{app:baseline}. For the evaluation of various diffusion models in unconditional generation (see \Cref{sec:uncond-eval}), we sample 50,000 sequences from each model. For the conditional generation across 15 species (see \Cref{sec:main-result}), we generate 4,000 sequences. In both cases, we use the DDPM sampler~\citep{song2020denoising} with 1,000 sequential denoising steps.

\subsubsection{Evaluating Diffusion Models on Mammalian Model Organisms}
\label{sec:uncond-eval}
\begin{table*}[ht]
    \centering
    \small
    \caption{Comparison of diffusion models on unconditional generation evaluated on EPD (256 bp) and EPD (2048 bp). Metrics include S-FID, \(\text{Cor}_{\text{TATA}}\), and \(\text{MSE}_{\text{TATA}}\). The \textbf{best} and \underline{second-best} scores are highlighted in bold and underlined, respectively.}

    \resizebox{0.9\linewidth}{!}{
    \begin{tabular}{
            l
            c
            c
            c
            c
            c
            c
            c
        }
        \toprule
                                          & \multicolumn{3}{c}{EPD (256 bp)} & \multicolumn{3}{c}{EPD (2048 bp)}                                                                                                                                               \\
        \cmidrule(lr){2-4} \cmidrule(lr){5-7}
        Model                             & {S-FID $\downarrow$}                 & $\text{Cor}_{\text{TATA}}\uparrow$   & { $\text{MSE}_{\text{TATA}}$ $\downarrow$}                  & {S-FID $\downarrow$}           & $\text{Cor}_{\text{TATA}}\uparrow$ & { $\text{MSE}_{\text{TATA}}$ $\downarrow$}                  \\
        \midrule
        Random (Reference)                & 119.0                                & -0.241                               & 8.21                            & 106.0                          & 0.030                              & 1.86                             \\
        Sample from Training Set          & 0.509                                & 1.0                                  & 0                             & 0.100                          & 0.999                              & 0                             \\
        \midrule
        VAE                               & 295.0                                & -0.167                               & 26.5                             & 250.0                          & 0.007                              & 9.40                             \\
        BitDiffusion                      & 405                                  & 0.058                                &5.29                             & 100.0                          & 0.066                              & 5.91                             \\
        D3PM (small)                      & \underline{97.4}                                 & 0.0964                               & 4.97 & \underline{94.5}                           & 0.363                              &  \textbf{1.50} \\
        D3PM (large)                      & 161.0                                & -0.208                               & \underline{4.75}                             & 224.0                          & 0.307                              & 8.49                             \\
        DDSM (Time Dilation)              & 504.0                                & \underline{0.897}                                & 13.4                             & 1113.0                         & \underline{0.839}                              & 2673.7                             \\
        DiscDiff (Ours)                   & \textbf{57.4}            & \textbf{0.973}           & \textbf{0.669}                            & \textbf{45.2}     & \textbf{0.858}         & \underline{1.74}                            \\
        \bottomrule
    \end{tabular}
    }
    \label{table:unconditional}
\end{table*}

One prerequisite of \textsf{Fast A\&E} (\Cref{alg:ae-fast}) is that the diffusion model $P^{DM}_\beta$ should be properly trained and provide accurate approximations of underlying data distribution. We first evaluate existing Diffusion Models on a subset of EPD datasets. This subset includes sequences from four mammalians \textit{H. Sapiens} (human), \textit{Rattus Norvegicus} (rat), \textit{Macaca mulatta}, and \textit{Mus musculus} (mouse), which collectively represent 50\% of the total dataset. Training on this subset allows for a more precise assessment of the generative algorithm's accuracy in a unconditional generation setting.

\textbf{Metrics}
\begin{enumerate}
    \item \textbf{Motif Distribution Correlation (\(\text{Cor}_{\text{M}}\))} and \textbf{Mean Square Error (\(\text{MSE}_{\text{M}}\))}: \(\text{Cor}_{\text{M}}\) is the Pearson correlation between the motif distributions of generated and natural DNA sequences for motifs like TATA-box, GC-box, Initiator, and CCAAT-box. \(\text{MSE}_{\text{M}}\) is the average squared differences between these motif distributions.
    
    \item \textbf{S-FID (Sei Fréchet Inception Distance)}: Measures the distance between distributions of generated and natural DNA sequences in latent space similar to the FID metric~\citep{heusel2017gans} for images, replacing the encoder with the pre-trained genomic neural network, Sei~\citep{zhou2015predicting}.
\end{enumerate}

The results are presented in \Cref{table:unconditional}, indicating that most existing models perform worse than the simple baseline DiscDiff proposed here, as measured by  S-FID, \(\text{Cor}_{\text{TATA}}\), and \(\text{MSE}_{\text{TATA}}\). TATA is one of the most fundamental motifs for gene transcription -- a special type of protein called transcription factors binds to TATA tokens on the DNA as shown in \Cref{fig:tata-demo}. It changes the shape of DNA and then enables the gene transcription. The failure of existing diffusion models to capture the TATA-box distribution indicates a potential gap in existing research. In the following, we hereby use DiscDiff as the $p^{DM}_\beta$ to initialize the \textsf{A\&E} algorithm.

\subsection{Multi-species DNA Sequences Generation}
\label{sec:main-result}
\begin{figure}
	\centering
	\includegraphics[width=0.9\linewidth]{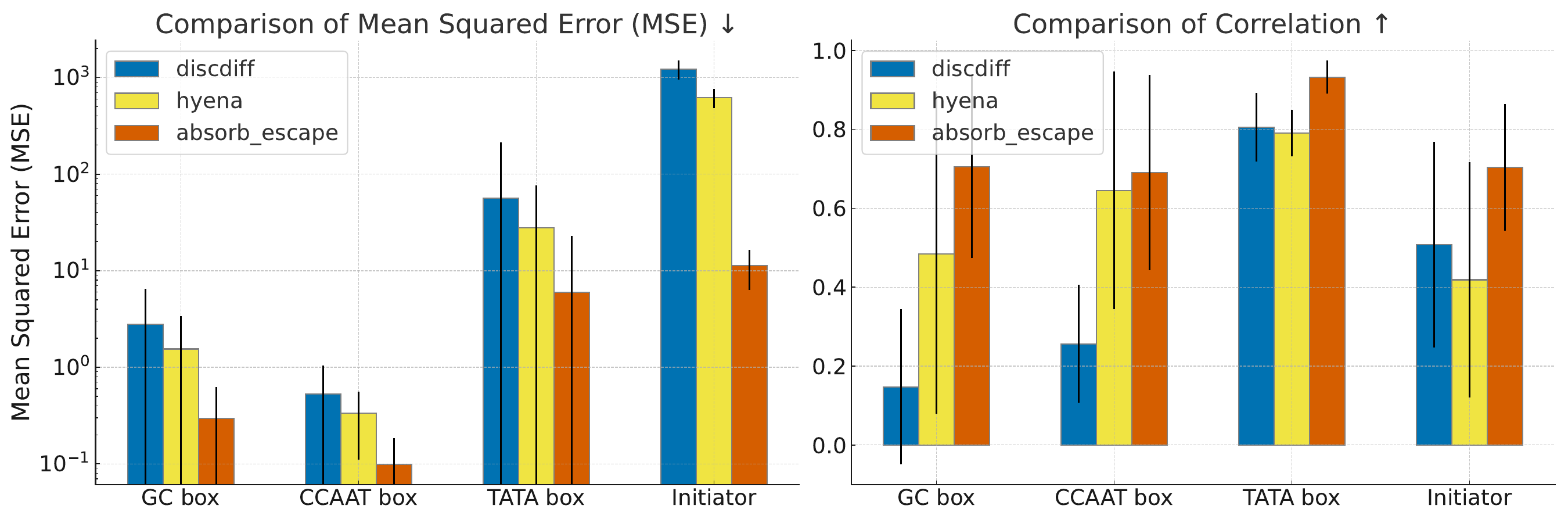}
	\caption{\textbf{The average MSE and Correlation between generated and natural DNA distributions for each model and motif type across 15 species.} \textsf{Fast A\&E} outperforms \textit{Hyena} and \textit{DiscDiff}, generating the most realistic sequences with the lowest MSE and highest Correlation across four motif types. This pattern is consistent across all 15 species.}
	\label{fig:mainresult-bar}
 \vspace{-4mm}
\end{figure}

We compare our model \textsf{Fast A\&E} with \textit{Hyena}~\citep{nguyen2024hyenadna} and the best-performing diffusion model from \Cref{sec:uncond-eval}, \textit{DiscDiff}, on the task of generating species-specific DNA sequences. 

\paragraph{Motif-centric Evaluation} We consider four types of motifs closely related to promoter activities $\{\text{TATA-box},\text{GC-box}, \text{Initiator},\text{CCAAT-box}\}$. We calculate 4 types of motif distributions for 15 species across 3 models, resulting in 180 frequency distributions.

\begin{wrapfigure}[18]{r}{0.53\linewidth}
	\centering
    \vspace{-5mm}
	\includegraphics[width=0.9\linewidth]{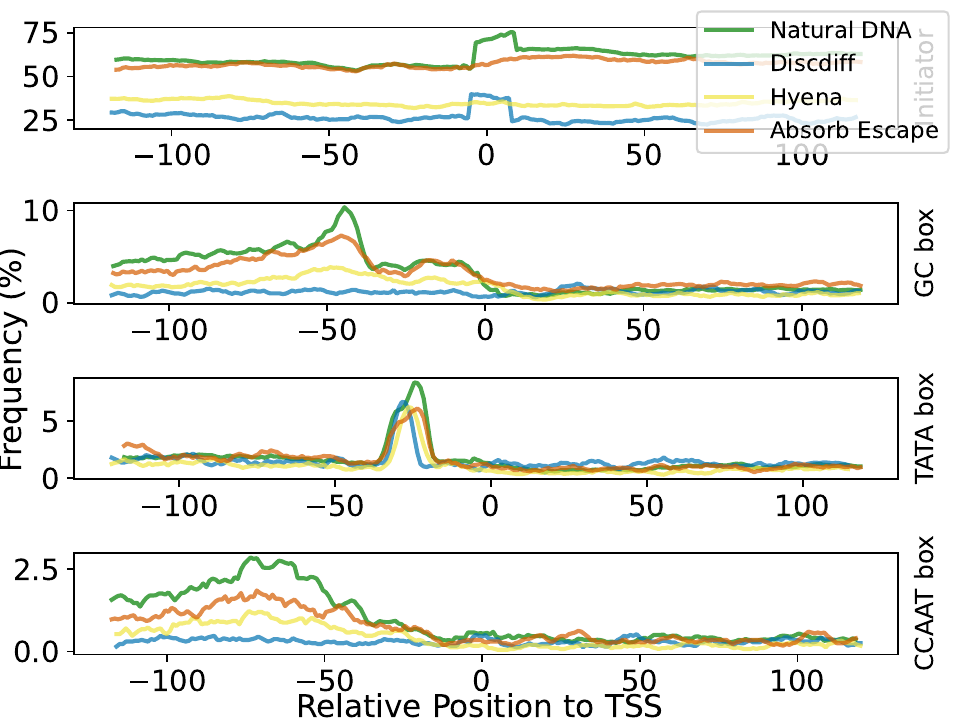}
	\caption{\textbf{Motif distributions in macaque DNA compared across natural DNA,  \textsf{FAST A\&E}, DiscDiff, and Hyena}. \textsf{FAST A\&E} closely aligns with natural DNA, especially for the TATA and GC motifs.}
	\label{fig:main-result-motif}
\end{wrapfigure}

\Cref{fig:mainresult-bar} shows the average MSE and Correlation between generated and natural DNA distributions for each model and motif type across 15 species. \textsf{Fast A\&E} improves upon \textit{Hyena} and \textit{DiscDiff}, generating the most realistic sequences across almost all species and motif types. It achieves the lowest MSE and highest Correlation across all four motifs. This pattern is consistent across all 15 species. The motif plots for all 15 species are provided in \Cref{app:appendix_motif_distributions}. As an example, \Cref{fig:main-result-motif} shows the motif distributions of sequences generated by the three models versus real DNA sequences for \textit{macaque}. \textsf{Fast A\&E} closely resembles the natural motif distribution, especially for the TATA and GC box motifs, while \textit{Hyena} and \textit{DiscDiff} fail to capture the values or trends accurately.

\textbf{Sequence Diversity}
 To assess the diversity of the generated sequences, we applied BLASTN~\citep{kent2002blat} to check (1) the similarity between the training set (Natural DNA) and generated sequences from three models, and (2) the similarity within the generated sequences. BLASTN takes a query DNA sequence and compares it with a database of sequences, returning all the aligned sequences in the database that are similar to the query. For each alignment, an alignment score, the alignment length (AlignLen), and statistical significance (eValue) are provided to indicate the quality of the alignment, where a larger alignment score, a smaller statistical significance (eValue), and a longer alignment sequence (AlignLen) indicate a higher similarity between the query sequence and the database sequences. Ideally, when using generated sequences to query the training dataset, a good generative sequence should align better than a random sequence, but not replicate the sequences in the training set.

\begin{wraptable}[13]{r}{0.55\textwidth}
    \vspace{-5mm}
    \small
    \caption{\textbf{BLASTN Results}}
    \label{table:blast}
    \vspace{4mm}
    \centering
    \setlength{\tabcolsep}{4pt} 
\centering
\begin{tabular}{@{}llll@{}}
\textbf{Query vs. Database} & \textbf{Score}$\uparrow$ & \textbf{eValue}$\downarrow$ & \textbf{AlignLen}$\uparrow$ \\
\toprule
Random vs. Natural DNA & 17.78 & 1.1769 & 24.2 \\ 
\midrule
A\&E vs. Natural & 21.39 & 0.1695 & 35.9 \\
Hyena vs. Natural & 22.89 & 0.2895 & 40.1 \\ 
DiscDiff vs. Natural & 20.25 & 0.2098 & 31.4 \\ 
\midrule
A\&E vs. A\&E & 20.14 & 0.0968 & 33.69 \\ 
Hyena vs. Hyena & 19.57 & 0.0843 & 28.7 \\ 
DiscDiff vs. DiscDiff & 20.95 & 0.1029 & 37.6 \\
Natural vs. Natural DNA & 57.06 & 0.0633 & 77.6 \\ 
\bottomrule
\end{tabular}
\end{wraptable}

\Cref{table:blast} shows the results of the BLASTN algorithm. From the table, \textit{DiscDiff}, \textit{Hyena}, and \textsf{A\&E} all satisfied the mentioned criteria. In terms of the diversity within the groups, none of the algorithms generated repetitive sequences. Furthermore, \textsf{A\&E} tends to lie between Hyena and DiscDiff in terms of the diversity of the generated sequences, implying that \textsf{A\&E} may combine the properties of AR and DM models. One notable fact is that the alignment scores are very high within the natural sequences, potentially indicating that natural sequences naturally have repetitive properties (conservative motifs), while the generative sequences do not have this characteristic.

\begin{figure}
	\centering
	\includegraphics[width=1\linewidth]{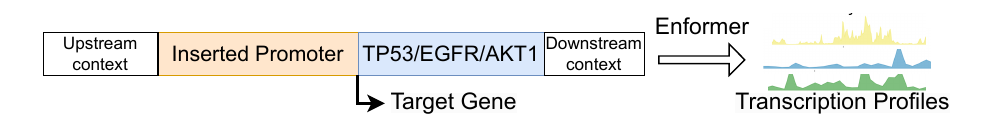}	\caption{\textbf{Evaluation of Generated Promoters for gene regulation through Genome Integration} }
	\label{fig:genome-integration}
 \vskip -0.3in
\end{figure}

\textbf{Genome Integration with Promoter Sequences}
As shown in \Cref{fig:genome-integration}, to evaluate the functional properties of sequences generated by \textit{Hyena}, \textit{DiscDiff}, and \textit{A\&E}, we inserted the generated promoter sequences of length 128 bp upstream (5') of three commonly studied genes in oncology: \textbf{TP53}, \textbf{EGFR}, and \textbf{AKT1}~\citep{olivier2010tp53,bethune2010epidermal,rudolph2016akt1}, which are closely related to tumor activities. Our goal was to determine which model generates promoter sequences that produce gene expression levels closest to those of natural promoters when reinserted into the human genome.

\begin{wraptable}[10]{r}{0.45\textwidth}
    \vspace{-5mm}
    \small
    \caption{\textbf{Sum of Squared Errors (SSE) of Transcription Profiles between Real and Generated Sequences}}
    \label{table:transcript-profile}
    \centering
    \setlength{\tabcolsep}{4pt} 
\centering
\begin{tabular}{@{}llll@{}}
\textbf{} & \textbf{TP53}$\downarrow$ & \textbf{EGFR}$\downarrow$ & \textbf{AKT1}$\downarrow$ \\
\toprule
Random & 278.18 & 8.09 & 65.70 \\ 
\midrule
A\&E  & \textbf{17.21} & \textbf{0.28} & \textbf{1.65} \\
Hyena & 36.25 & 0.89 & 2.88 \\ 
DiscDiff & 124.03 & 2.17 & 25.50 \\ 
\bottomrule
\end{tabular}
\end{wraptable}

We use Enformer~\citep{avsec2021effective} to predict transcription profiles. Enformer takes a DNA sequence of 200k bps as input and outputs a matrix \( P \in \mathbb{R}^{896 \times 638} \), representing a multi-cell type transcription profile. We sampled 300 promoter sequences from each source: \textit{Natural DNA promoters}, \textit{Hyena}, \textit{DiscDiff}, and \textsf{A\&E}. For each set, we calculated the average transcription profile across the sequences. The Sum of Squared Errors (SSE) between these average transcription profiles of the generated sequences and those of natural promoters are shown in \Cref{table:transcript-profile}. The results indicate that \textsf{A\&E} produces the smallest SSE, suggesting it best captures the properties of natural DNA. This finding highlights the potential of generative algorithms to create promoter sequences that effectively regulate gene expression, with applications in bioproduct manufacturing and gene therapy.

\subsubsection{Sensitivity Analysis of $T_{\text{Absorb}}$} 
\begin{figure}[ht]
    \begin{center}
    \centerline{\includegraphics[width=0.45\columnwidth]{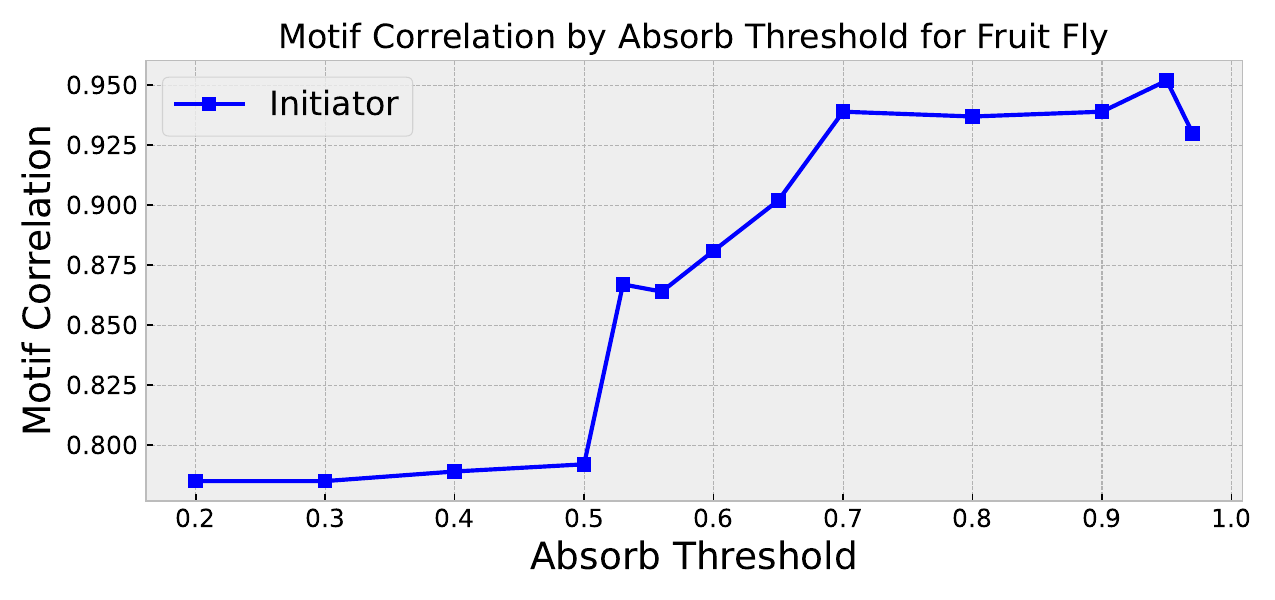}}
    \caption{\textbf{Sensitivity of A\&E under various \( T_{\text{absorb}} \)}. For each value of \( T_{\text{absorb}} \), 3000 sequences are generated and compared with natural DNA. The correlation between the generated sequences and natural DNA increases initially as \( T_{\text{absorb}} \) increases, and then it flattens. An optimal \( T_{\text{absorb}} \) can be selected based on the validation set, or a default value of 0.85 can be used.}
    \label{fig:Absorb-Escape-trade-off}
    \end{center}
    \vskip -0.4in
\end{figure}

We perform a \textit{sensitivity analysis} of A\&E algorithm over hyperparameter \( T_{\text{absorb}} \). As shown in \Cref{fig:Absorb-Escape-trade-off}, with a small \( T_{\text{absorb}} \), the sequences generated by A\&E are dominated by the diffusion model. As \( T_{\text{absorb}} \) increases, the AR helps to correct the errors made by the DM. Finally, when \( T_{\text{absorb}} \) is larger than 0.7, the correlation flattens and fluctuates. In conclusion, A\&E is robust under different values of \( T_{\text{absorb}} \), and it is best to use the validation dataset to choose the optimal value. However, a wide range of \( T_{\text{absorb}} \) can still be used with improved performance.

%% file: sections/conclusion.tex
\section{Conclusion}
This paper demonstrates that (i) both the AutoRegressive (AR) model and Diffusion Models (DMs) fail to accurately model DNA sequences due to the heterogeneous nature of DNA sequences when used separately, and (ii) this limitation can be overcome by introducing \textsf{A\&E}, a novel sampling algorithm that combines AR models and DMs. Additionally, we developed a fast implementation of the proposed algorithm, \textsf{Fast A\&E}, which enables efficient generation of realistic DNA sequences without the repetitive function evaluations required by conventional sampling algorithms. Experimental results across 15 species show that \textsf{Fast A\&E} consistently outperforms single models in generating DNA sequences with functional and structural similarities to natural DNA, as evidenced by metrics such as Motif Distribution, Sequence Diversity, and Genome Integration. Regarding the future work, the generated DNA sequences still require validation through wet-lab experiments before they can be directly used in clinical settings. 

%% file: Appendix/appendix.tex
\appendix

\section{A simple baseline latent diffusion model for discrete data: DiscDiff}
\label{app:discdiff}

\begin{figure*}[!t]
    \begin{center}
    \centerline{\includegraphics[width=\textwidth]{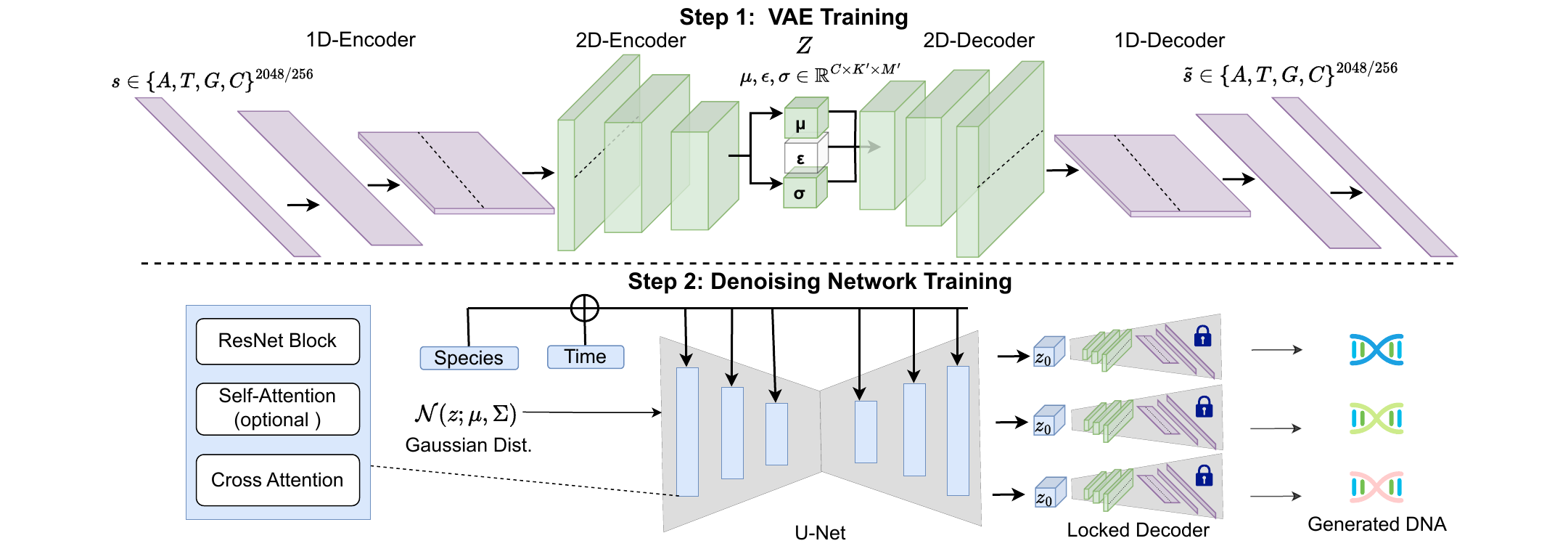}}

\caption{DiscDiff Model: A two-step process for DNA sequence generation. \textbf{Step 1: VAE Training}: 
A sequence $s \in \{A, T, G, C\}^{2048/256}$ is encoded via a 1D-Encoder to a 2D-Encoder. The latent space representation $Z$ with parameters $\mu, \epsilon, \sigma$ is then decoded back to $\tilde{s}$ through a 2D-Decoder and 1D-Decoder.
\textbf{Step 2: Denoising Network Training}: 
The latent representation $Z$ is processed through a denoising network comprising a ResNet Block, optional Self-Attention, and Cross Attention, with species and time information. The network outputs a Gaussian distribution $N(z; \mu, \Sigma)$. A U-Net architecture takes this distribution to produce various $z_0$ representations, which a Locked Decoder (fronzen parameters) used to generate the final DNA sequences.}
        \label{pic:discdiff}
    \end{center}
\end{figure*}

\subsection{The DiscDiff Model}

We design DiscDiff, a Latent Discrete Diffusion model for DNA generation tasks.
As shown in \Cref{pic:discdiff}, this model is structured into two main components: a Variational-Auto-Encoder (VAE) and a denoising model. The VAE consists of an encoder $\mathbf{E}: \b s \mapsto \b z$, which maps discrete input sequence $\b s$ to a continuous latent variable $\b z$, and a decoder $\mathbf{D}: \b z \mapsto \tilde{\b s}$, which reverts $\b z$ back to $\tilde{s}$ in the discrete space.
The denoising model $\mathbf{\mathbf\varepsilon_{\mathbf\theta}}(\mathbf z_t,t)$ is trained to predict the added noise $\mathbf\varepsilon$ in the latent space.

\subsubsection{VAE Architecture}

The choice of VAE architecture in LDMs is critical and often domain-specific. We find that mapping the input data to a higher dimension space can help to learn a better denoising network, generating more realistic DNA sequences. We hereby propose to use a two-stage VAE architecture as shown in \Cref{pic:discdiff}.

The first stage encoder $\mathbf{E_{\mathbf\phi}}_1: \mathbb{N}^L_4 \rightarrow \mathbb{R}^{K \times M}
$ maps $\b s \in \mathbb{N}^L_4$ to a 2D latent space $\b z_1 \in \mathbb{R}^{K \times M}$, where $K$ is the number of channels and $M$ is the length of the latent representation.
The second stage encoder $\mathbf{E_{\mathbf\phi}}_2 : \mathbb{R}^{1 \times K \times M} \rightarrow \mathbb{R}^{ C \times K' \times M'}$ first adds a dummy dimension to $\b z_1$ such that $\mathbf z_1 \in \mathbb{R}^{1 \times K \times M}$ and then maps it to 3d latent space $\mathbf z \in \mathbb{R}^{ C \times K' \times M'}$,
where $C$ is the number of channels, $K'$ and $M'$ are the reduced dimensions of $K$ and $M$ respectively.
The decoder in the first and second stage are $\mathbf{D_{\mathbf\theta}}_1$ and $\mathbf{D_{\mathbf\theta}}_2$ respectively. Which are symmetric to the encoders. Overall,
we have $\mathbf z = \mathbf{E_{\mathbf\phi}}(\mathbf s) = \mathbf{E_{\mathbf\phi}}_2(\mathbf{E_{\mathbf\phi}}_1(\mathbf s))$, and the reconstruction is $\tilde{\mathbf s} = \mathbf{D_{\mathbf\theta}}(\mathbf z) = \mathbf{D_{\mathbf\theta}}_1(\mathbf{D_{\mathbf\theta}}_2(\mathbf z))$.

\subsubsection{VAE Loss} 

When training the VAE, we propose to use Cross Entropy (CE) as reconstruction loss. The loss function is given by:

\begin{align*}
    \mathbf{L}_{\theta,\phi} = \underbrace{
        \mathbb{E}_{p(\mathbf s)}\left[\mathbb{E}_{q_{\mathbf \phi}(\mathbf z|\mathbf s)}\left[
            -\sum_{l=1}^{L}\sum_{i=1}^{4}\delta_{is_l} \log p_{\mathbf\theta}(\mathbf s_l|\mathbf z)
    \right]\right]}_{\text{Reconstruction Loss}} + \\
    \beta\cdot \underbrace{\mathbb{E}_{p(\mathbf s)}\left[
            \mathrm{KL}(q_{\mathbf\phi}(\mathbf z|\mathbf s) \,||\, \mathcal{N}(\mathbf z; \mathbf \mu, \mathbf \Sigma))
            \right]}_{\text{KL Divergence}}
\end{align*}

where $\delta_{ij}$ is the Kronecker delta, \( p_{\mathbf\theta}(\mathbf s|\mathbf z) \) is the probabilistic decoder output from $\mathbf{D_{\mathbf\theta}}$; \( q_{\mathbf\phi}(\mathbf z|\mathbf s) \) is the probabilistic output from encoder $\mathbf{E_{\mathbf\phi}}$ that represents the approximate posterior of the latent variable \(\mathbf z \) given the input \( \mathbf s \); $\mathcal{N}(\mathbf z;\mathbf \mu, \mathbf\Sigma)$ is the prior on \( \mathbf z \). Here we use a simple isotropic. $\beta$ is a mixing hyperparameter. $\beta$ is set to $10^{-5}$ in the experiments used in this paper. 

\subsubsection{Denoising Network Training}
Once $\mathbf{D_{\mathbf\theta}}$ and $\mathbf{E_{\mathbf\phi}}$ are trained in the first step,
we train a noise prediction $\mathbf\varepsilon_{\mathbf\theta}$ in the latent space $\mathbf z = \mathbf{E_{\mathbf\phi}}(\mathbf s)$ with
\Cref{eq:denoising-loss}. 
\begin{equation}
    \mathbb{E}_{\mathbf z,t \sim U[1,T],\mathbf\varepsilon \sim \mathcal{N}(\mathbf 0,\mathbf I)} \left[  \|\mathbf\varepsilon -\mathbf{\mathbf\varepsilon_{\mathbf\theta}}(\mathbf z_t,t)  \|^2_2 \right]
    \label{eq:denoising-loss}
\end{equation}

\section{Toy Experiment Training Details}
\label{app:toy-exp-setup}

We train both HyenaDNA and DiscDiff on an NVIDIA A100-PCIE-40GB using the Adam optimizer. For HyenaDNA, the learning rate is set to 0.0001, and we use the model \texttt{heyenadna-large-1m-seqlen} for this task. The maximum number of epochs is set to 100, with early stopping enabled to facilitate early convergence.

For DiscDiff, the VAE is trained with a learning rate of 0.0001, while the UNet is trained with a learning rate of 0.00005. DiscDiff is trained for 600 epoches; during the inference time, we use DDPM~\cite{ho2020denoising} sampler with 1000 denoising steps.

\section{Convergence Proof of the Absorb \& Escape Algorithm}
\label{app:proof-ab-convergence}
In this section, we provide a proof for the convergence of the Absorb \& Escape (\textsf{A\&E}) algorithm under certain assumptions. The convergence proof will demonstrate that the sequence generated by the \textsf{A\&E} algorithm converges to a sample from the target distribution \(p^{C}_{\theta,\beta}(\mathbf{x})\).

\subsection{Assumptions}
We make the following assumptions for the convergence proof:
\begin{enumerate}
    \item The autoregressive model \(p^{AR}_\theta(\mathbf{x})\) and the diffusion model \(p^{DM}_\beta(\mathbf{x})\) are both properly trained and provide accurate approximations of the underlying data distribution.
    \item The initial sample \(\mathbf{x}^0 \sim p^{DM}_\beta(\mathbf{x})\) is a valid sample from the diffusion model.
    \item The segments \(\{\mathbf{s}_1, \mathbf{s}_2, \ldots, \mathbf{s}_n\}\) of the sequence \(\mathbf{x}\) are chosen such that each segment is homogeneous.
    \item The subset of segments \(\mathcal{S}\) is chosen randomly and includes all segments over multiple iterations.
    \item The conditional distribution \(p(\mathbf{x}_{i:j}|\mathbf{x}_{0:i-1}, \mathbf{x}_{j+1:L})\) approximated by \(p^{AR}_\theta(\mathbf{x}_{i:j}|\mathbf{x}_{0:i-1})\) is accurate.
\end{enumerate}

\subsection{Proof}
We aim to show that the \textsf{A\&E} algorithm produces samples from the target distribution \(p^{C}_{\theta,\beta}(\mathbf{x})\). We do this by showing that the Markov chain defined by the \textsf{A\&E} algorithm has \(p^{C}_{\theta,\beta}(\mathbf{x})\) as its stationary distribution.

\paragraph{Step 1: Initialization}
The initial sample \(\mathbf{x}^0 \sim p^{DM}_\beta(\mathbf{x})\) is drawn from the diffusion model. This ensures that \(\mathbf{x}^0\) is a valid sample from \(p^{DM}_\beta(\mathbf{x})\).

\paragraph{Step 2: Absorb Step}
For each segment \(\mathbf{s}_k\) in the subset \(\mathcal{S}\), the Absorb step samples \(\tilde{\mathbf{x}}_{i:j}'\) from the conditional distribution \(p(\mathbf{x}_{i:j}|\mathbf{x}_{0:i-1}, \mathbf{x}_{j+1:L})\). Under the assumption that \(p^{AR}_\theta(\mathbf{x}_{i:j}|\mathbf{x}_{0:i-1})\) accurately approximates this conditional distribution, the refined segment \(\tilde{\mathbf{x}}_{i:j}'\) is a valid sample from the target conditional distribution.

\paragraph{Step 3: Escape Step}
The Escape step updates the segment \(\tilde{\mathbf{x}}_{i:j}^t\) to \(\tilde{\mathbf{x}}_{i:j}'\). This ensures that the updated sequence \(\tilde{\mathbf{x}}^t\) incorporates the refinement from the autoregressive model.

\paragraph{Step 4: Stationary Distribution}
To show that the Markov chain defined by the \textsf{A\&E} algorithm converges to \(p^{C}_{\theta,\beta}(\mathbf{x})\), we need to show that \(p^{C}_{\theta,\beta}(\mathbf{x})\) is the stationary distribution of this Markov chain.

The transition probability for the \textsf{A\&E} algorithm is given by the product of the probabilities of the Absorb and Escape steps:
\[
P(\mathbf{x} \rightarrow \mathbf{x}') = P_{\text{Absorb}}(\mathbf{x} \rightarrow \mathbf{x}') P_{\text{Escape}}(\mathbf{x}' \rightarrow \mathbf{x}).
\]

Given that \(p^{DM}_\beta(\mathbf{x})\) captures the overall structure and \(p^{AR}_\theta(\mathbf{x})\) refines the segments, the composed distribution \(p^{C}_{\theta,\beta}(\mathbf{x})\) is achieved by iteratively applying the Absorb and Escape steps. We need to show that the stationary distribution satisfies:
\[
\int p^{C}_{\theta,\beta}(\mathbf{x}) P(\mathbf{x} \rightarrow \mathbf{x}') d\mathbf{x} = p^{C}_{\theta,\beta}(\mathbf{x}').
\]

Using the detailed balance condition for the Markov chain:
\[
p^{C}_{\theta,\beta}(\mathbf{x}) P(\mathbf{x} \rightarrow \mathbf{x}') = p^{C}_{\theta,\beta}(\mathbf{x}') P(\mathbf{x}' \rightarrow \mathbf{x}).
\]

Given that the Markov chain is ergodic, the detailed balance condition implies that \(p^{C}_{\theta,\beta}(\mathbf{x})\) is the stationary distribution.

\paragraph{Step 5: Ergodicity}
Ergodicity ensures that the Markov chain will visit all possible states given sufficient iterations. The random selection of segments \(\mathcal{S}\) and the iterative updates in the \textsf{A\&E} algorithm guarantee that all parts of the sequence are refined over time.

\paragraph{Conclusion}
By satisfying the detailed balance condition and ergodicity, we have shown that the Markov chain defined by the \textsf{A\&E} algorithm converges to the target distribution \(p^{C}_{\theta,\beta}(\mathbf{x})\). Therefore, the \textsf{A\&E} algorithm produces samples from the composed distribution \(p^{C}_{\theta,\beta}(\mathbf{x})\) as the number of iterations \(t\) approaches infinity.

\section{Experiment Details}
\label{app:baseline}

\paragraph{Baselines} The details about the \textbf{architecture and implementation} of the baseline models are as below:
\begin{itemize}
\item DNADiffusion~\citep{pinello2024dna}: We enhance the current DNADiffusion implementation for DNA synthesis, originally from the DNA-Diffusion project\footnote{\url{https://github.com/pinellolab/DNA-Diffusion}}, by expanding the models to encompass 380 million parameters. This network is composed of Convolutional Neural Networks (CNNs), interspersed with layers of cross-attention and self-attention. The learning rate is set to $0.0001$.

\item DDSM~\citep{avdeyev2023dirichlet}: We scale up the original implementation of the denoising network used for promoter design in DDSM\footnote{\url{https://github.com/jzhoulab/ddsm}} to what is the corresponding size of the network given 470 million parameters. It is a convolution-based architecture with dilated convolution layers. The learning rate is set to $0.00001$.
\item D3PM~\citep{austin2021structured}: We take the implementation of D3PM for biological sequence generation from EvoDiff~\citep{alamdari2023protein}\footnote{\url{https://github.com/microsoft/evodiff}}, adopting the algorithm for DNA generation. We use the original implementation of the denoising network, which has two versions: with sizes of 38M and 640M. We hereby have D3PM (small) and D3PM (big), respectively. The learning rate for both D3PM (small) and D3PM (large) are set to $0.0001$.
\item Hyena~\citep{nguyen2024hyenadna}: We modify the RegLM~\citep{lal2024reglm}\footnote{\url{https://github.com/Genentech/regLM}}, a existing work uses hyena for DNA generation. Four pretrained Hyena models of different sizes (hyenadna-large-1m-seqlen, hyenadna-medium-160k0seqlen, heynadna-small-32k-seqlen, and hyenaana-tiny-16k-seqlen-d128) are downloaded from HuggingFace\footnote{\url{https://huggingface.co/LongSafari}} and used for full-size fine-tuning, we apply the fine-tuned models for generations on EPD-GenDNA. The learning rate for fine-tuning is set to $0.0001$.

\item DiscDiff: A 2D-UNet of 500 Million parameters are used as the denoising network. See \Cref{app:discdiff} for the implementation details. The learning rate is set to $0.00005$ for UNet training, $0.0001$ for VAE training. 
    
\end{itemize}

For the \textsf{Fast A\&B} algorithm, we set the $T_{absorb}$ to 0.80.

\section{Content List of Supplementary Code and Data}
\label{app:code}
Our code folder includes the following sub-folders, within each folder there is a readme file, detailing the steps to run the code. The below is the list of sub-folders.
\subsection{Toy Experiment}
Include the source code to reproduce \Cref{sec:toy-example}. No external package is required to run the code except for python

\subsection{DiscDiff}
 A implementation of the discdiff baseline in \Cref{app:discdiff}. Please follow the readme for running the code.

\subsection{AbsorbEscape}

This includes an implementation of the proposed algorithm in \Cref{sec:ae-implementation}; however, it  depends on the external AR models, please adopt it accordingly to AR models which you want to try.

\subsection{EPD Data}
We include the training dataset used for producing the main results, which includes 160K DNA sequences from EPD, each sequence has a length of 256 bp.

\section{Motif Distributions for 15 species}
\label{app:appendix_motif_distributions}
We plot TATA-box, GC content, Initiator, and CCAAT-box for 15 speces as below.

\begin{figure}[h]
    \centering
    \includegraphics[width=0.8\textwidth]{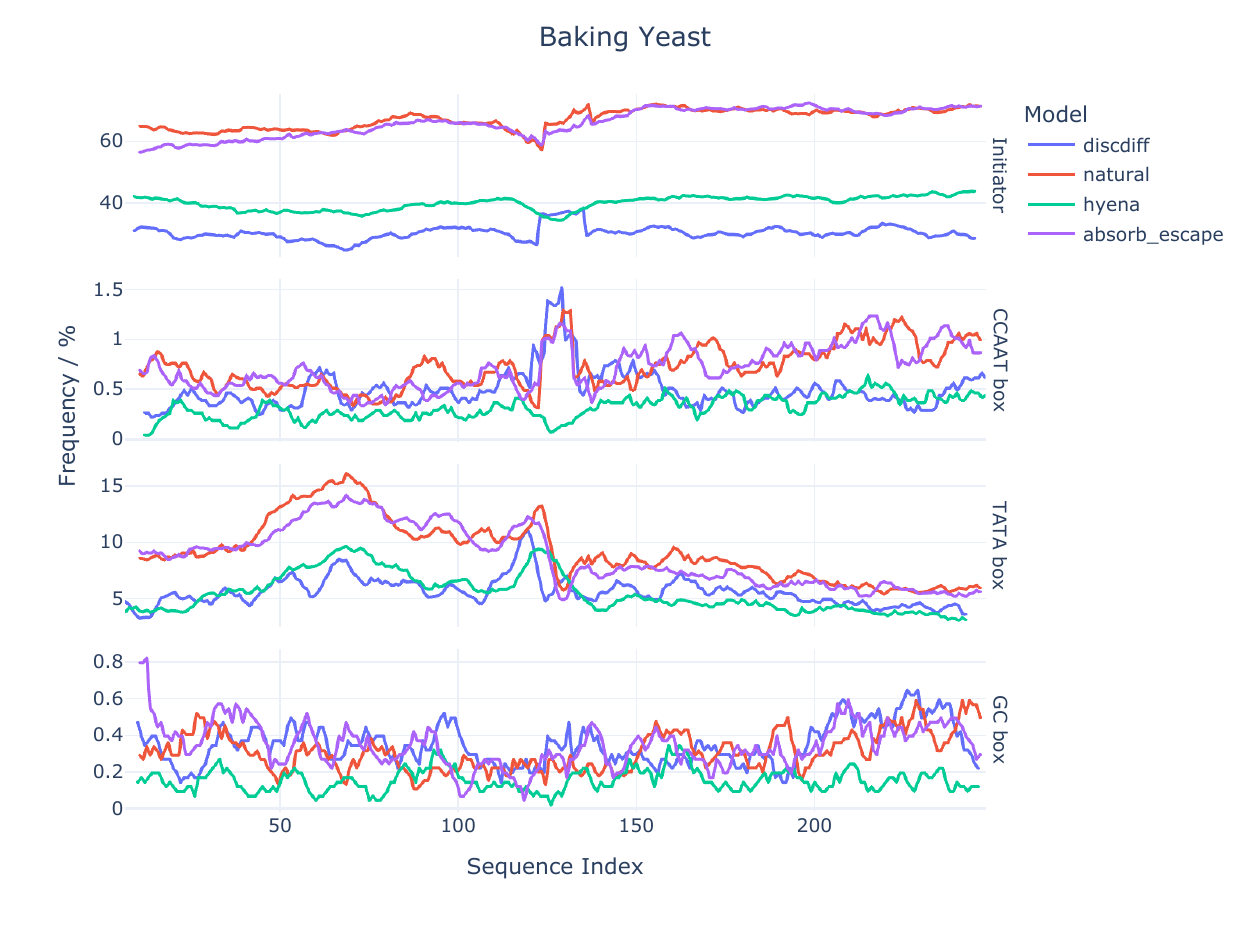}
    \caption{Baking Yeast}
    \label{fig:motif-distributions1}
\end{figure}

\begin{figure}[h]
    \centering
    \includegraphics[width=0.8\textwidth]{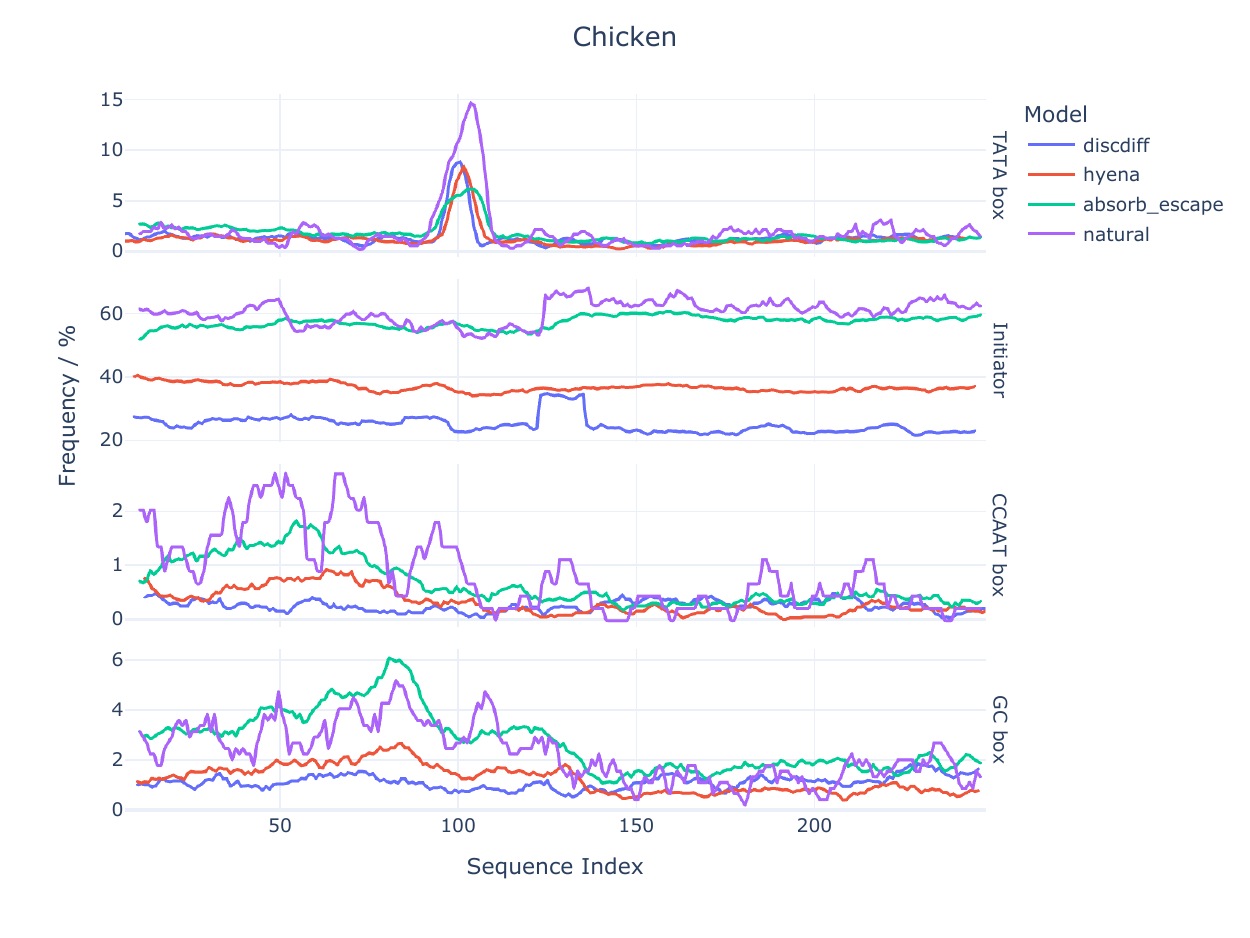}
    \caption{Chicken}
    \label{fig:motif-distributions2}
\end{figure}

\begin{figure}[h]
    \centering
    \includegraphics[width=0.8\textwidth]{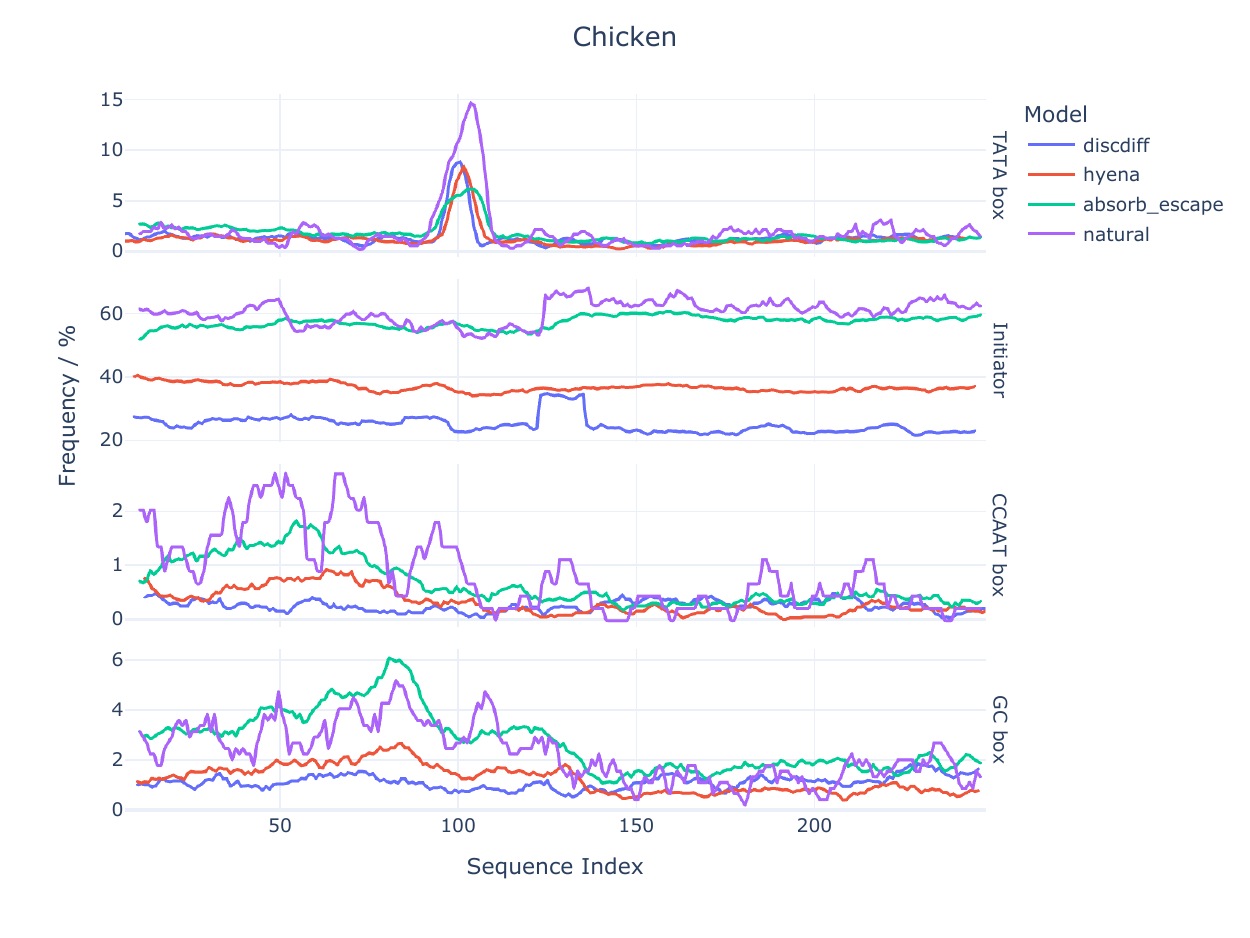}
    \caption{Chicken}
    \label{fig:motif-distributions3}
\end{figure}

\begin{figure}[h]
    \centering
    \includegraphics[width=0.8\textwidth]{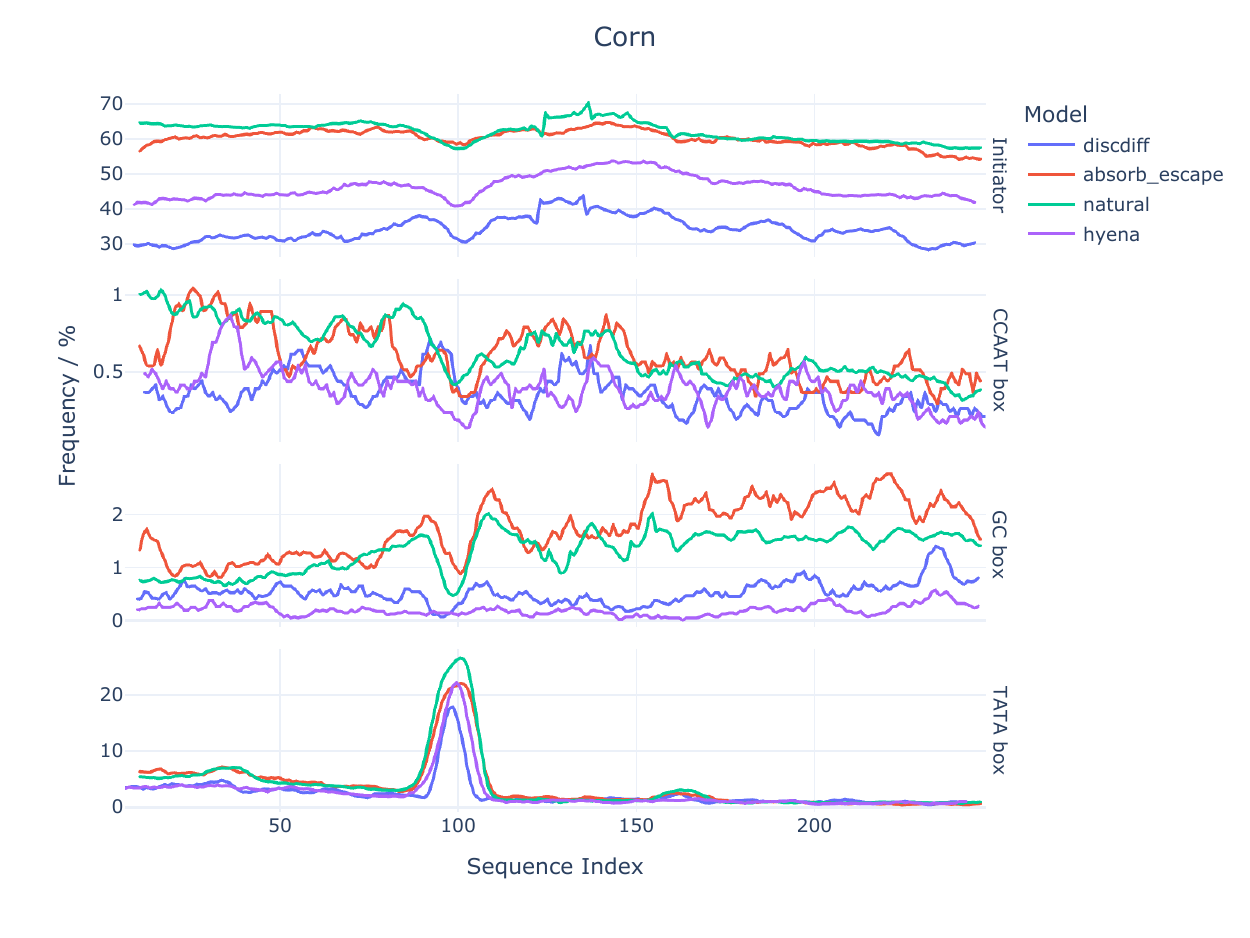}
    \caption{Corn}
    \label{fig:motif-distributions4}
\end{figure}

\begin{figure}[h]
    \centering
    \includegraphics[width=0.8\textwidth]{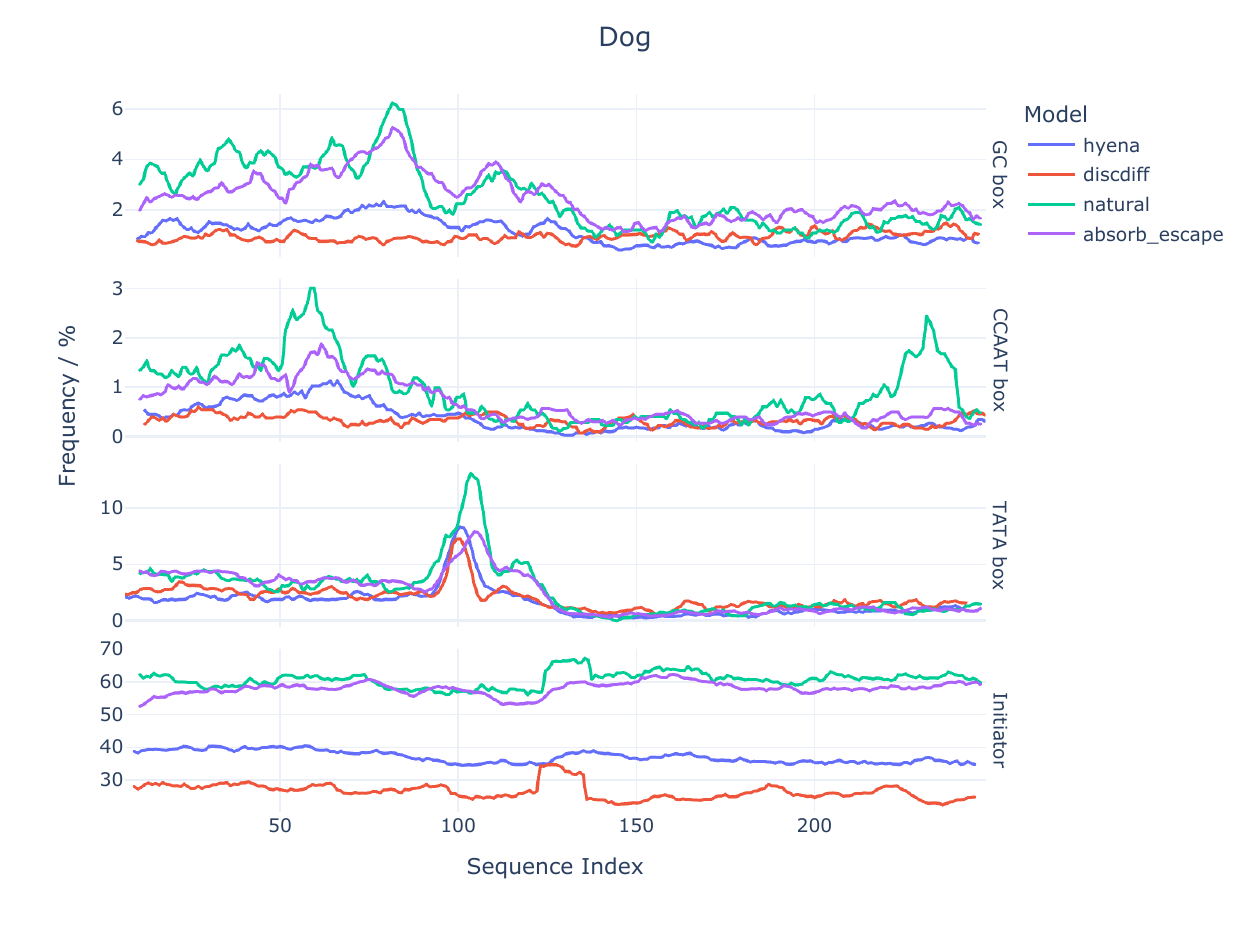}
    \caption{dog}
    \label{fig:motif-distributions5}
\end{figure}

\begin{figure}[h]
    \centering
    \includegraphics[width=0.8\textwidth]{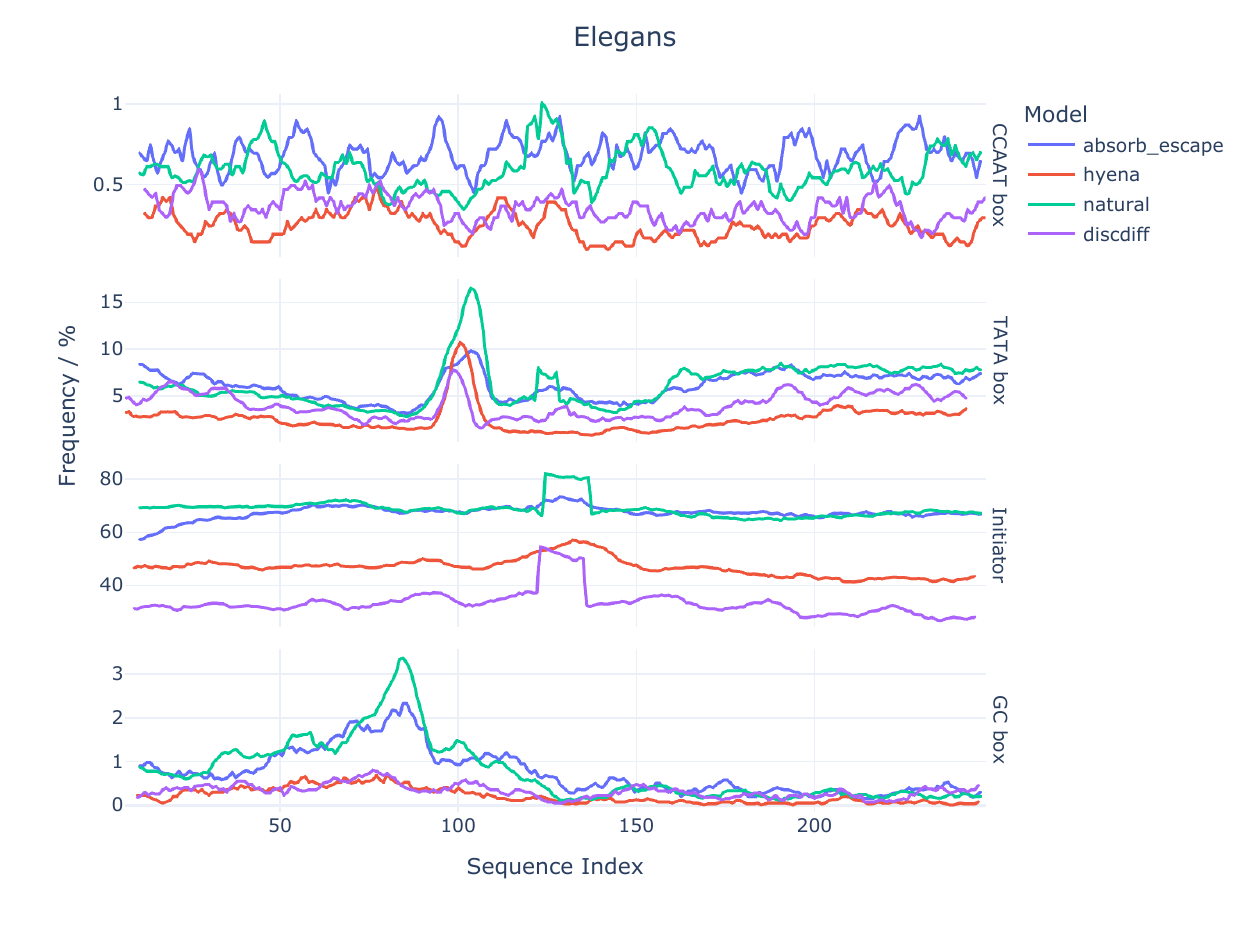}
    \caption{elegans}
    \label{fig:motif-distributions6}
\end{figure}

\begin{figure}[h]
    \centering
    \includegraphics[width=0.8\textwidth]{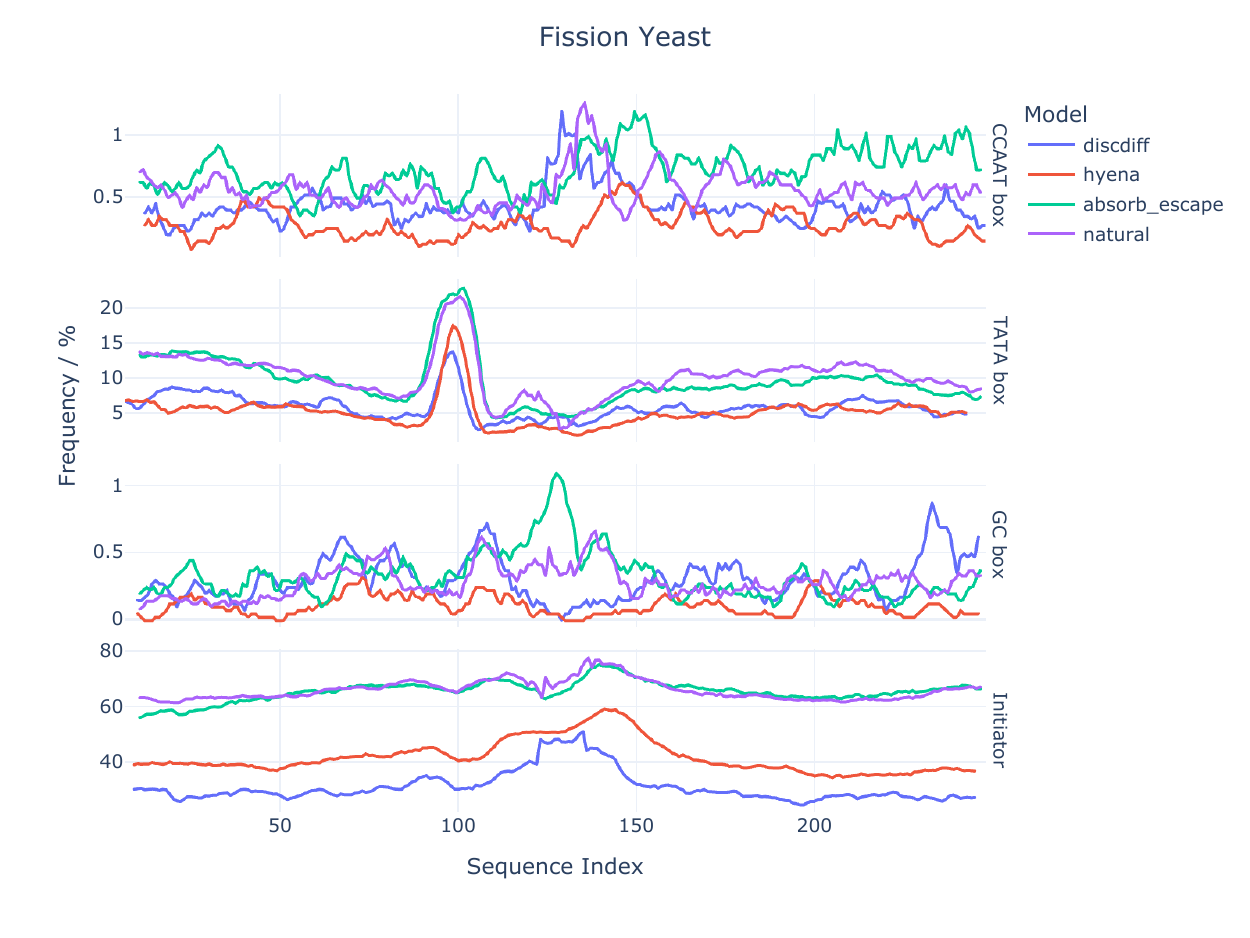}
    \caption{fission yeast}
    \label{fig:motif-distributions7}
\end{figure}

\begin{figure}[h]
    \centering
    \includegraphics[width=0.8\textwidth]{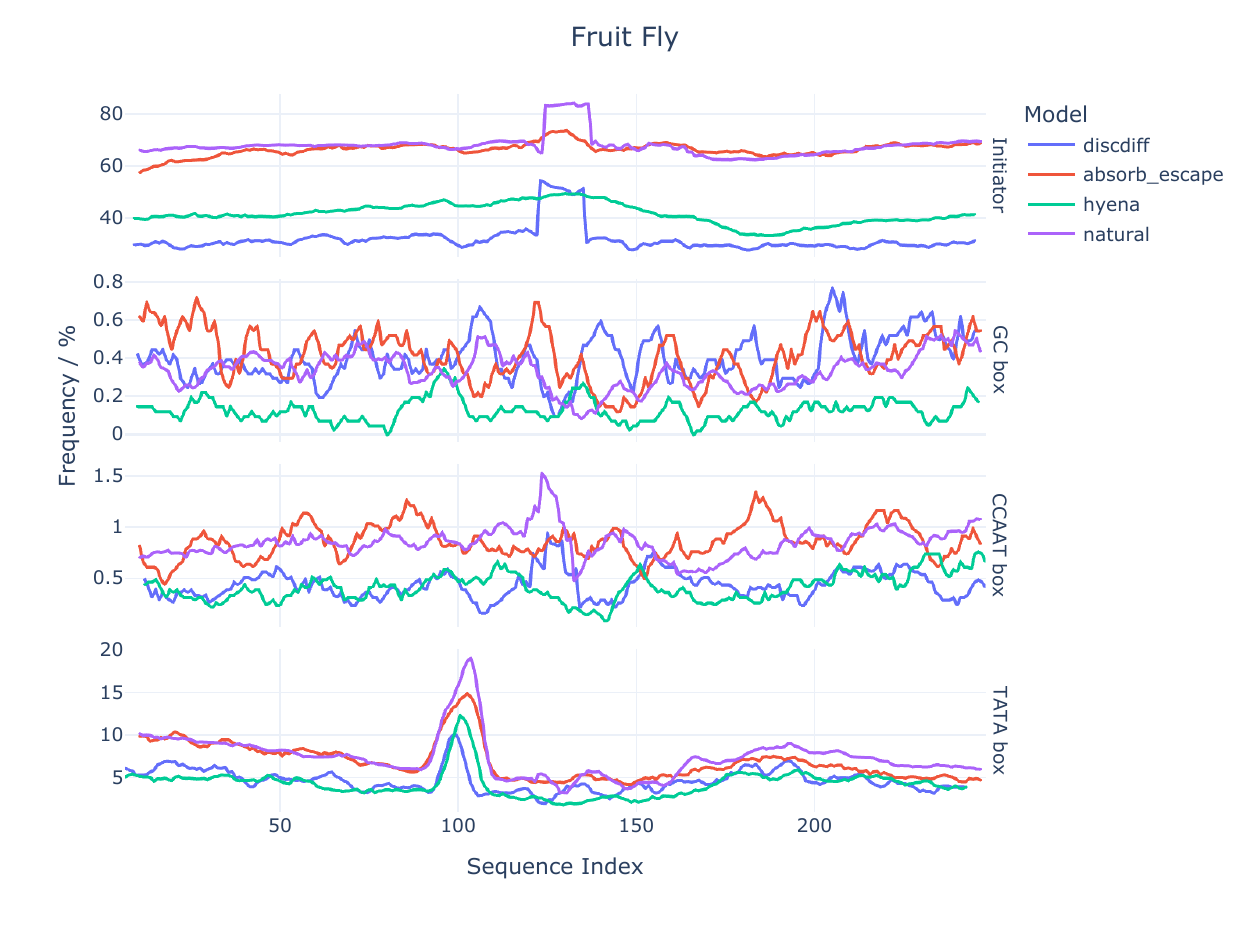}
    \caption{fruit fly}
    \label{fig:motif-distributions8}
\end{figure}

\begin{figure}[h]
    \centering
    \includegraphics[width=0.8\textwidth]{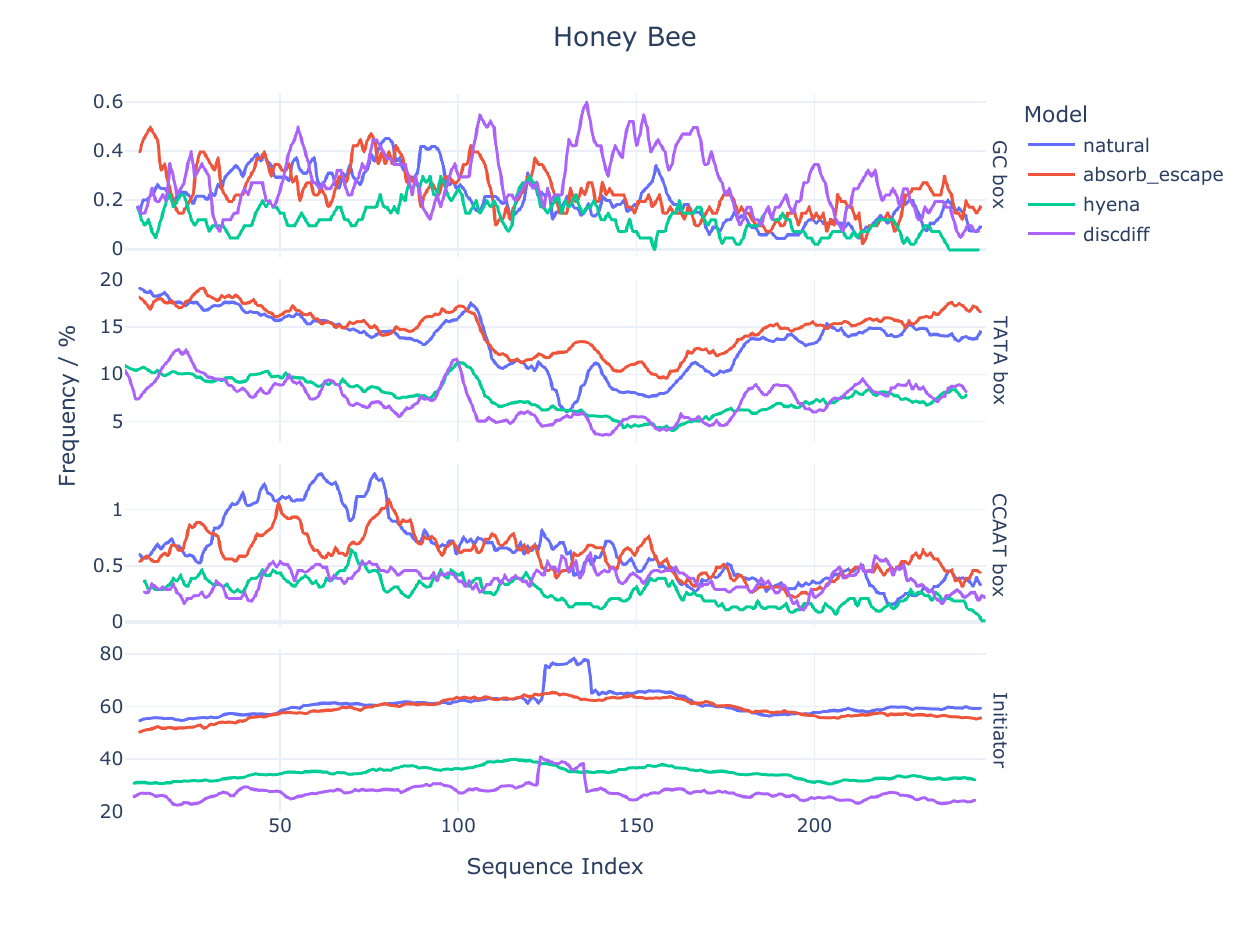}
    \caption{honey bee}
    \label{fig:motif-distributions9}
\end{figure}
\begin{figure}[h]
    \centering
    \includegraphics[width=0.8\textwidth]{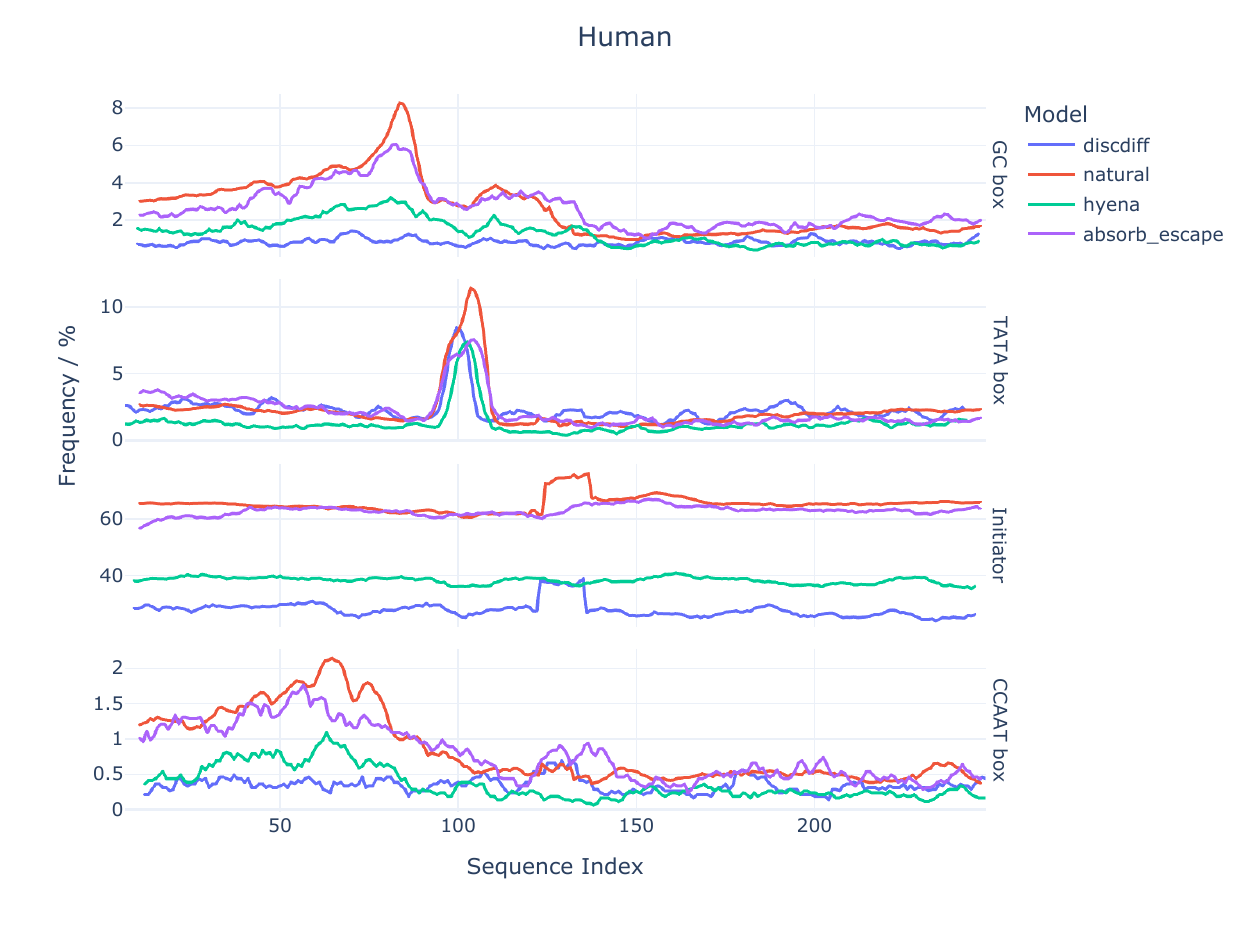}
    \caption{human}
    \label{fig:motif-distributions10}
\end{figure}
\begin{figure}[h]
    \centering
    \includegraphics[width=0.8\textwidth]{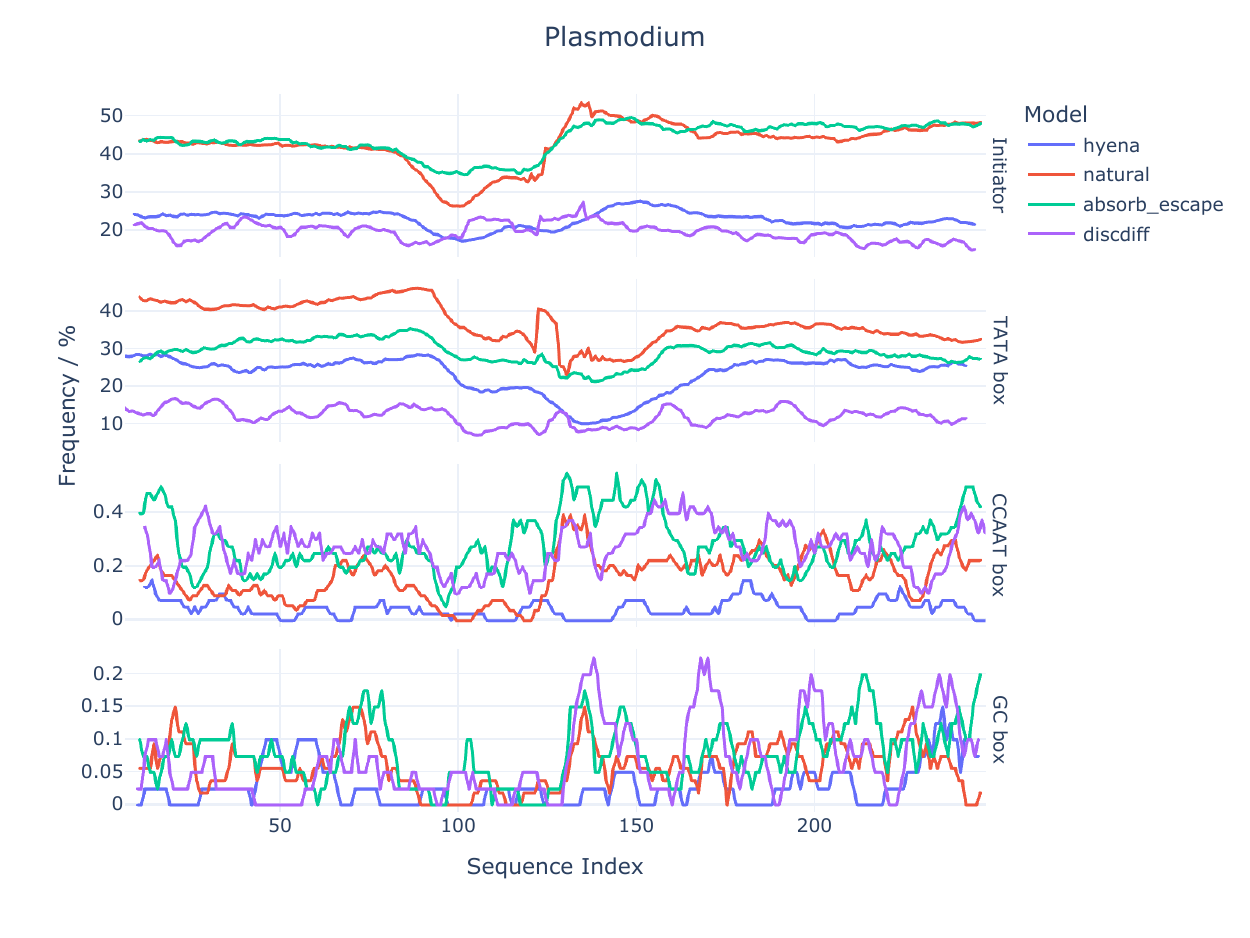}
    \caption{plasmodium}
    \label{fig:motif-distributions11}
\end{figure}
\begin{figure}[h]
    \centering
    \includegraphics[width=0.8\textwidth]{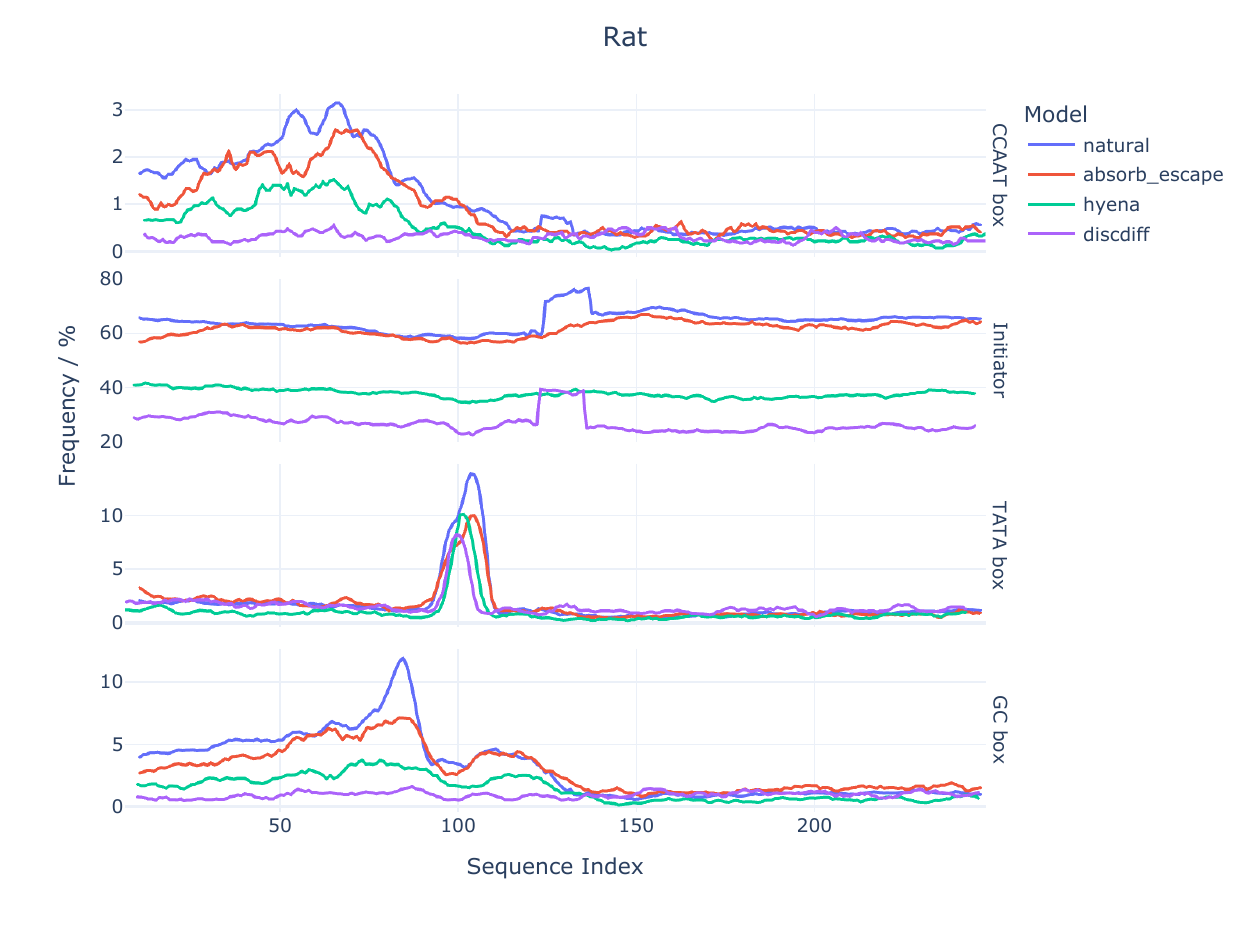}
    \caption{rat}
    \label{fig:motif-distributions12}
\end{figure}
\begin{figure}[h]
    \centering
    \includegraphics[width=0.8\textwidth]{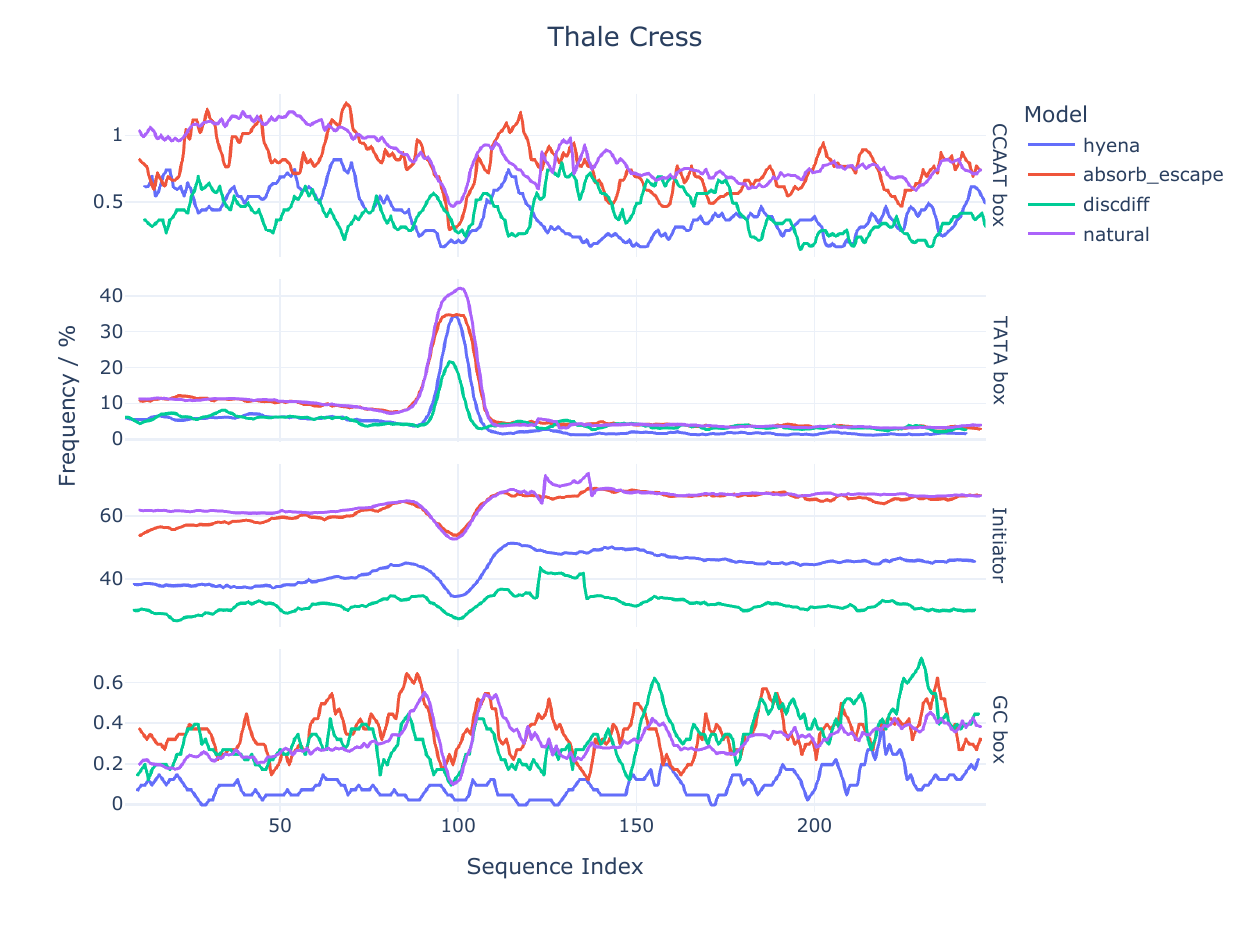}
    \caption{thale cress}
    \label{fig:motif-distributions13}
\end{figure}
\begin{figure}[h]
    \centering
    \includegraphics[width=0.8\textwidth]{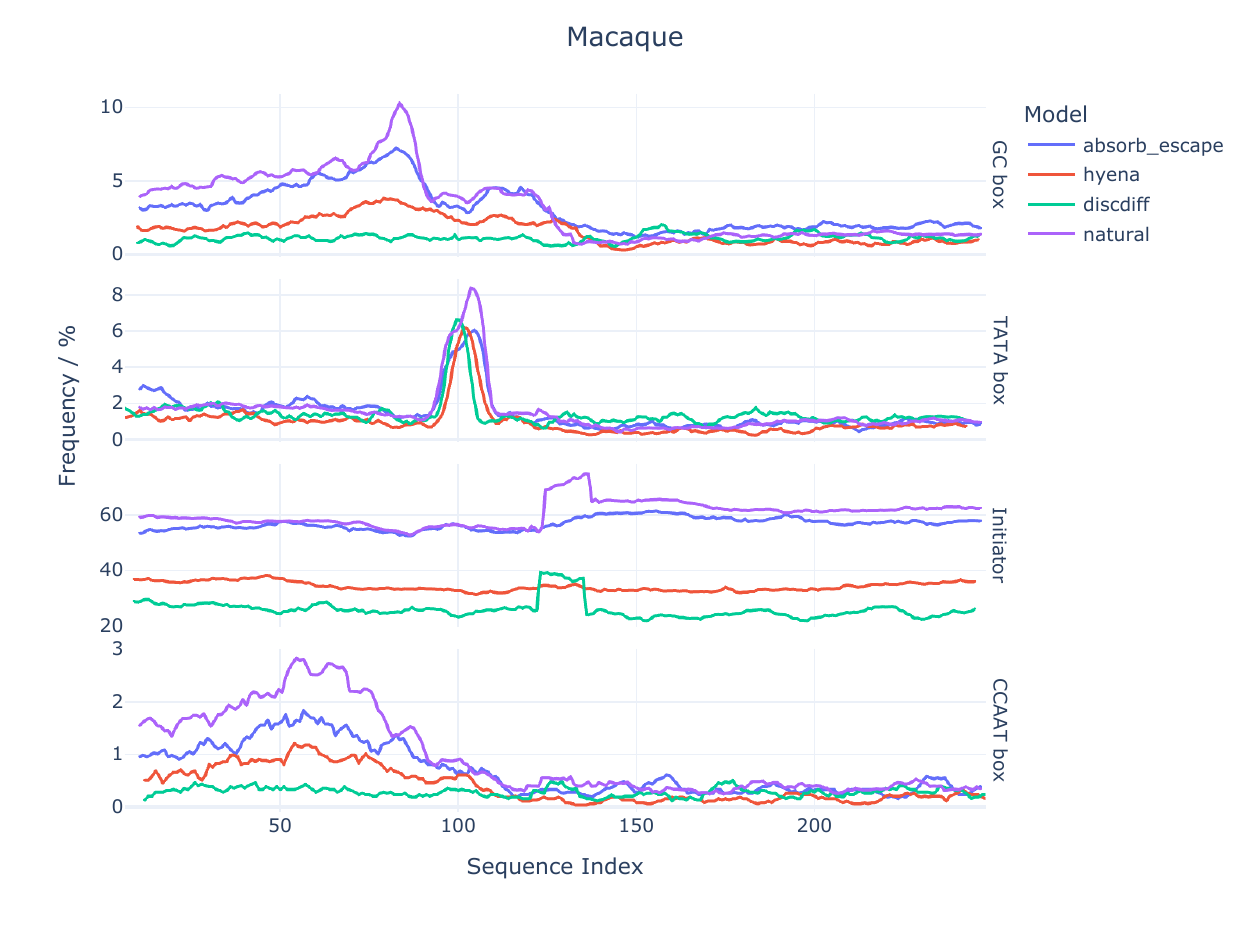}
    \caption{macaque}
    \label{fig:motif-distributions14}
\end{figure}

\begin{figure}[h]
    \centering
    \includegraphics[width=0.8\textwidth]{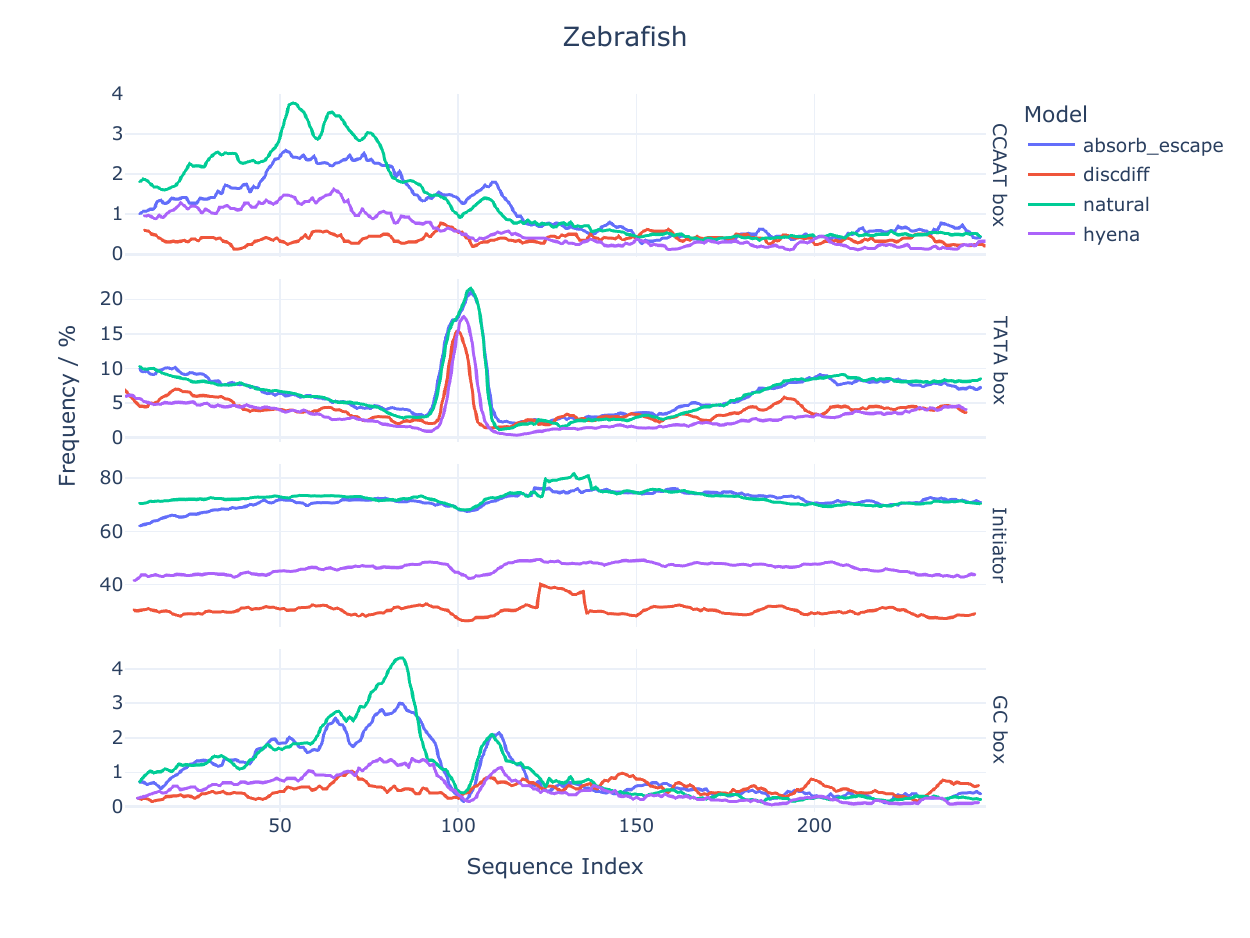}
    \caption{ zebrafish}
    \label{fig:motif-distributions15}
\end{figure}

%% file: sections/checklist.tex
\section*{NeurIPS Paper Checklist}


\begin{enumerate}

\item {\bf Claims}
    \item[] Question: Do the main claims made in the abstract and introduction accurately reflect the paper's contributions and scope?
    \item[] Answer: \answerYes{} 
    \item[] Justification: The sections related to the contributions are listed at the end of the introduction.

\item {\bf Limitations}
    \item[] Question: Does the paper discuss the limitations of the work performed by the authors?
    \item[] Answer: \answerYes{}
    \item[] Justification: We talked about the assumption required by the algorithm in the \Cref{app:proof-ab-convergence}, we also talked about other limitations through the paper.

\item {\bf Theory Assumptions and Proofs}
    \item[] Question: For each theoretical result, does the paper provide the full set of assumptions and a complete (and correct) proof?
    \item[] Answer: \answerYes{} 
    \item[] Justification: Can be found in \Cref{app:proof-ab-convergence}

    \item {\bf Experimental Result Reproducibility}
    \item[] Question: Does the paper fully disclose all the information needed to reproduce the main experimental results of the paper to the extent that it affects the main claims and/or conclusions of the paper (regardless of whether the code and data are provided or not)?
    \item[] Answer: \answerYes{} 
    \item[] Justification: For the Toy Example experiment details see \Cref{app:toy-exp-setup}, for the baseline and main result set up see \Cref{app:baseline}

\item {\bf Open access to data and code}
    \item[] Question: Does the paper provide open access to the data and code, with sufficient instructions to faithfully reproduce the main experimental results, as described in supplemental material?
    \item[] Answer: \answerYes{} 
    \item[] Justification: See \Cref{app:code} for the instructions to access the code.

\item {\bf Experimental Setting/Details}
    \item[] Question: Does the paper specify all the training and test details (e.g., data splits, hyperparameters, how they were chosen, type of optimizer, etc.) necessary to understand the results?
    \item[] Answer: \answerYes{} 
    \item[] Justification: See \Cref{sec:main-result} and \Cref{app:baseline}.

\item {\bf Experiment Statistical Significance}
    \item[] Question: Does the paper report error bars suitably and correctly defined or other appropriate information about the statistical significance of the experiments?
    \item[] Answer: \answerYes{} 
    \item[] Justification: For the main result which compares the motif distributions across 15 species, we compute the error bar. For some of the other tasks requiring heavy compute, we didn't repeat the experiment. 

\item {\bf Experiments Compute Resources}
    \item[] Question: For each experiment, does the paper provide sufficient information on the computer resources (type of compute workers, memory, time of execution) needed to reproduce the experiments?
    \item[] Answer: \answerYes{}{} 
    \item[] Justification: See \Cref{app:baseline} and \Cref{sec:main-result}
    
\item {\bf Code Of Ethics}
    \item[] Question: Does the research conducted in the paper conform, in every respect, with the NeurIPS Code of Ethics \url{https://neurips.cc/public/EthicsGuidelines}?
    \item[] Answer: \answerYes{} 
    \item[] Justification: 

\item {\bf Broader Impacts}
    \item[] Question: Does the paper discuss both potential positive societal impacts and negative societal impacts of the work performed?
    \item[] Answer:  \answerYes{} 
    \item[] Justification:  Mainly in the introduction part
    
\item {\bf Safeguards}
    \item[] Question: Does the paper describe safeguards that have been put in place for responsible release of data or models that have a high risk for misuse (e.g., pretrained language models, image generators, or scraped datasets)?
    \item[] Answer:  \answerNA{}
    \item[] Justification: 

\item {\bf Licenses for existing assets}
    \item[] Question: Are the creators or original owners of assets (e.g., code, data, models), used in the paper, properly credited and are the license and terms of use explicitly mentioned and properly respected?
    \item[] Answer: \answerYes{} 
    \item[] Justification: All the code repos - The MIT License (MIT)
     eukaryotic promoter database - CC-BY 4.0

\item {\bf New Assets}
    \item[] Question: Are new assets introduced in the paper well documented and is the documentation provided alongside the assets?
    \item[] Answer: \answerYes{}{} 
    \item[] Justification: See \Cref{app:code}

\item {\bf Crowdsourcing and Research with Human Subjects}
    \item[] Question: For crowdsourcing experiments and research with human subjects, does the paper include the full text of instructions given to participants and screenshots, if applicable, as well as details about compensation (if any)? 
    \item[] Answer: \answerNA{}{} 
    \item[] Justification: No Crowdsourcing Involved

\item {\bf Institutional Review Board (IRB) Approvals or Equivalent for Research with Human Subjects}
    \item[] Question: Does the paper describe potential risks incurred by study participants, whether such risks were disclosed to the subjects, and whether Institutional Review Board (IRB) approvals (or an equivalent approval/review based on the requirements of your country or institution) were obtained?
    \item[] Answer: \answerNA{} 
    \item[] Justification: No Crowdsourcing Involved

\end{enumerate}